# A mathematical model of COVID-19 with an underlying health condition using fraction-order derivative

Samuel Okyere[1*,] Joseph Ackora-Prah[1,] Ebenezer Bonyah[2] and Mary Osei Fokuo[3]

[1]Department of Mathematics, Kwame Nkrumah University of Science and Technology, Ghana

[3]Department of Mathematics Education, Akenten Appiah-Menka University of Skills Training and Enterpreneurial Development, Kumasi, Ghana

[3]Department of Mathematics, St. Monica's College of Education, Ghana

Corresponding Author: Samuel Okyere; okyere2015@gmail.com

## Abstract

Studies have shown that some people with underlying conditions such as cancer, heart failure, diabetes and hypertension are more likely to get COVID-19 and have worse outcomes. In this paper, a fractional-order derivative is proposed to study the transmission dynamics of COVID-19 taking into consideration population having an underlying condition. The fractional derivative is defined in the Atangana – Beleanu – Caputo (ABC) sense. For the proposed model, we find the basic reproductive number, the equilibrium points and determine the stability of these equilibrium points. The existence and the uniqueness of the solution are established along with Hyers –Ulam Stability. The numerical scheme for the operator was carried out to obtain a numerical simulation to support the analytical results. COVID-19 cases from March to June 2020 of Ghana were used to validate the model. The numerical simulation revealed a decline in infections as the fractional operator was increased from 0.6 within the 120 days. Time-dependent optimal control was incorporated into the model. The numerical simulation of the optimal control revealed, vaccination reduces the number of individuals susceptible to the COVID-19, exposed to the COVID-19 and Covid-19 patients with and without an underlying health condition.

## 1 Introduction

It is now evident that people with underlying conditions such as cancer, cerebrovascular disease, heart failure, coronary artery disease, cardiomyopathies, diabetes mellitus, tuberculosis, and hypertension get severely ill from COVID-19 [1]. Studies have shown that some people with the above disabilities are more likely to get COVID-19 and have worse outcomes [2 – 4]. The elderly and those with underlying conditions are the most vulnerable groups for COVID -19 infections. It is estimated that 22% of the world population has at least one underlying health condition listed above that puts them at higher risk for severe COVID-19 disease [6]. Also, 4% of the world's population would be hospitalized if infected with COVID-19 [6].

Research conducted by Choi (2021) [28], on the mortality rates of 566,602 patients with coronavirus disease (COVID-19) in a Mexican community revealed the mortality rate of patients with the underlying health conditions being 12% and was four (4) times higher than that of patients without the underlying condition. Several studies have confirmed that COVID-19 is more severe in older people and those with the underlying condition of diabetes, lung or heart diseases [27].

COVID-19 has been a pandemic since the outbreak in China, with over 470 million cases reported throughout the world with over 6 million deaths. Africa alone has over 11 million confirmed cases of COVID-19 with 252,301 deaths as of 24th March 2022. Ghana reported its first COVID-19 confirmed case on the 12th of March 2020 and now has reported over 160,000 confirmed cases of COVID-19 since the outbreak [5]. Ghana has most of the underlying conditions mentioned that put the country at high risk of the COVID-19. According to the United Nations, 5.27 million people in Ghana has hypertension [29], between 3.3 and 6% of the population has diabetes [30 - 32]. Also according to the World Health Organisation data published in 2018, coronary heart disease deaths in Ghana reached 18,029 or 9.00% of total deaths [33].

There have been several preventive mechanisms adopted by various countries around the globe to curb the spread of COVID-19. Among them is the wearing of a mask, the use of sanitisers and lockdown to restrict the movement of people. Vaccines are also being recommended for use in a lot of countries. Other preventive methods adopted by the affected countries are; social distancing, hand washing with soap and water and ventilating indoor spaces. All these preventive methods have greatly helped in the fight against COVID-19 [9].

Existing studies on COVID-19 using mathematical models have been centered on the transmission dynamics in a population without taking into consideration the underlying health conditions of the patients [42]. Mathematical models have been proposed to study infectious diseases using fractional derivatives [14 – 19]. The fractional-order models are more precise and reliable compared to the integer-order models due to the enhanced degrees of freedom. The memory property allows knowledge in the past to be added in predictions [15, 38]. In recent times several models have been formulated in a fractional-order derivative. One of the most effective and reliable fractional operators is the Atangana -- Beleanu and Caputo (ABC) [14].

Authors in [14] formulated and analyzed a mathematical model of Syphilis with an emphasis on treatment in the sense of Caputo-Fabrizio (CF) and Atangana-Baleanu (Mittag-Leffler law) derivatives. Authors in [15] proposed an optimal control fractional model to study tuberculosis infection with the inclusion of diabetes resistance strain. Authors in [16] proposed a fractional model for the dynamics of tuberculosis infection using Caputo-Fabrizio derivative. In [25], the authors proposed a fractional-order differential equation model in the Caputo sense to study the HIV-1 infection of CD4+ T-cells with the effect of drug therapy.

Some authors have also proposed fractional-order derivatives to study the transmission of the COVID -19. In [35], the authors formulated a fractional epidemic model in the Caputo sense with the consideration of quarantine, isolation, and environmental impacts to examine the dynamics of the COVID-19 outbreak in Pakistan. In [12], the authors proposed a fractional-order derivative to study the transmission dynamics of COVID-19 in Wuhan city, China. In [27], the authors proposed a susceptible-asymptomatic--symptomatic -- recovered - deceased (SEIRD) model using the fractional-order derivative to study the spread of COVID-19 in Italy. In [36], the authors proposed a mathematical model of the coronavirus disease 2019 (COVID-19) to investigate the transmission and control mechanism of the disease in the community of Nigeria using the Atangana-Beleanu operator. In this work, we examine the work of [15, 35] to formulate a fractional-order model for the transmission of COVID-19 in a population with an underlying condition. To the best of our knowledge, modelling of COVID-19 in a population with an underlying condition using fractional-order derivative has never been explored.

## 2 Model Formulation

In this section, we examine the transmission dynamics of the COVID-19 in a population with an underlying condition through a mathematical model agreeing to the components of the viral transmittal in [15, 35]. We study the transmission of COVID-19 with a fraction of the population having an underlying condition by using fractional derivatives. The fractional derivative is defined in the ABC sense. We define the fractional-order operator as $\psi$ , where $0 \prec \psi \leq 1$. The population is partitioned into six (6) classes namely: Susceptible Individuals (S), Individuals with an underlying condition $(S_U)$, Individuals exposed to COVID-19 (E), Individuals with COVID-19 without an underlying condition $(C)$, COVID-19 patients with an underlying condition $(C_U)$, Vaccinated individuals (V) and Individuals removed from COVID-19 (R). Therefore, the total population, N is $N = S + S_U + E + C + C_U + V + R$

We assume that the population is homogeneously mixed, with no restriction of age, mobility or other social factors. All newborns are susceptible (no inherited immunity). We also assumed that those vaccinated don't contribute to the spread of the COVID-19 virus and once infected, you become exposed to the disease before becoming infectious. Individuals are recruited into the susceptible population at the rate $\Omega$ and they develop the underlying condition at rate $\lambda$. The Susceptible may be infected when they interact with those in class C or $C_U$. Then the infected person becomes exposed to the disease and hence move to class E at rate $\beta$. The proportion of the exposed that joined the C class is $\alpha\varphi^{\psi} E$ and the remaining joined the class $C_U$ at the rate $(1-\alpha)\varphi^{\psi}$. The parameter $\mu^{\psi}, \delta^{\psi}, \delta_1^{\psi}$ and $\delta_2^{\psi}$ are the natural death rate, COVID-19 disease-induced death, deaths COVID-19 and diabetes and deaths due to diabetes. The rate at which co-infected people recovers from COVID-19 is $\gamma_1$ and the rate at which COVID-19 only patient recovers is $\gamma^{\psi}$. The parameter $\eta^{\psi}$ is the vaccination rate. The flowchart of the model is given in Fig. 1.

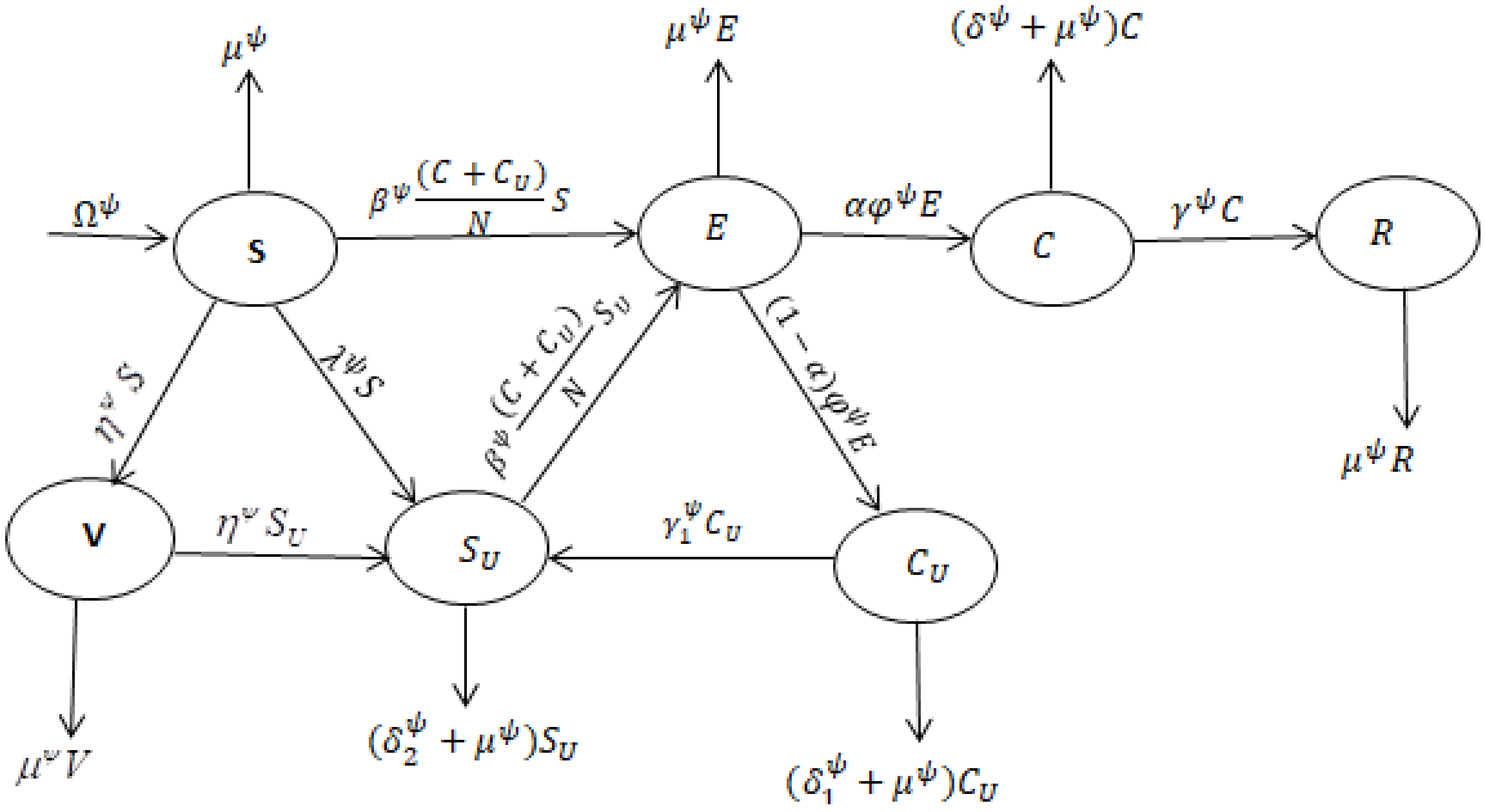

Fig. 1: Flowchart of the model

The fractional-order derivative describing the model is given as

$$
\begin{aligned}
{}^{ABC}D^{\psi}_{0,t}[S(t)] &= \Omega^{\psi} - \beta^{\psi}\left(\frac{C+C_U}{N}\right)S - (\mu^{\psi}+\lambda^{\psi})S,\\
{}^{ABC}D^{\psi}_{0,t}[S_U(t)] &= \lambda^{\psi}S + \gamma_1^{\psi}C_U - \beta^{\psi}\left(\frac{C+C_U}{N}\right)S_U - (\delta_2^{\psi}+\mu^{\psi})S_U,\\
{}^{ABC}D^{\psi}_{0,t}[E(t)] &= \beta^{\psi}\left(\frac{C+C_U}{N}\right)S + \beta^{\psi}\left(\frac{C+C_U}{N}\right)S_U - (\varphi^{\psi}+\mu^{\psi})E, \qquad (1)\\
{}^{ABC}D^{\psi}_{0,t}[C(t)] &= \alpha\varphi^{\psi}E - (\delta^{\psi}+\mu^{\psi}+\gamma^{\psi})C,\\
{}^{ABC}D^{\psi}_{0,t}[C_U(t)] &= (1-\alpha)\varphi^{\psi}E - (\delta_1^{\psi}+\mu^{\psi}+\gamma_1^{\psi})C_U,\\
{}^{ABC}D^{\psi}_{0,t}[V(t)] &= \eta^{\psi}(S+S_U) - \mu^{\psi}V,\\
{}^{ABC}D^{\psi}_{0,t}[R(t)] &= \gamma^{\psi}C - \mu^{\psi}R.
\end{aligned}
$$

With initial conditions

$$S(0)=S_0, S_U(0)=U_0, E(0)=E_0, C(0)=C_0, C_U(0)=C_{U0}, V(0)=V_0, R(0)=R_0.$$

### 2.1 Preliminaries

This section consists of the relevant definition of Atanga – Baleanu derivative and integration in Caputo sense taken from [18]

**Definition 1**

The Liouville- Caputo (LC) definition of the fractional derivative of order $\psi$ as defined in [19] is

$$ {}_{C}^{AB}D_{0,t}^{\psi} r(t) = \frac{1}{\Gamma(1-\psi)} \int_0^t (t-p)^{-\psi} r(p)dp, \qquad 0 \prec \psi \le 1 \qquad (2)$$

**Definition 2**

The Atangana – Baleanu definition in Liouville- Caputo sense as in, [18] is

$$ {}_{C}^{AB}D_{0,t}^{\psi} r(t) = \frac{B(\psi)}{\Gamma(1-\psi)} \int_0^t (t-p)^{-\psi} r(p)dp, \qquad (3)$$

where $B(\psi)=1-\psi+\frac{\psi}{\Gamma(\psi)}$, is the normalized function.

**Definition 3**

The corresponding fractional integral Atangana – Baleanu – Caputo derivative is defined as in [18] is

$$ {}_{C}^{AB}I_{0,t}^{\psi} r(t) = \frac{(1-\psi)}{B(\psi)} r(t) + \frac{\psi}{B(\psi)\Gamma(\psi)} \int_0^t (t-p)^{\psi-1} r(p)dp. \qquad (4)$$

They found that when $\psi=0$, they recovered the initial function and when $\psi=1$, the ordinary integral is obtained. The Laplace transform of equation (4) gives

$$ {}^{ABC}D_{0,t}^{\psi} r(t) = \frac{(1-\psi)}{B(\psi)} g(t) + \frac{B(\psi)G(q)q^{\psi} - q^{\psi-1}r(0)}{(1-\psi)\left(q^{\psi}+\frac{\psi}{1-\psi}\right)}. \qquad (5)$$

**Theorem 1**

For a function $r \in C[a,b]$, the following results hold [20]:

$$\left\| {}^{ABC}D_{0,t}^{\psi} r(t) \right\| \prec \frac{B(\psi)}{(1-\psi)} \left\| r(t) \right\|, \text{ where } \left\| r(t) \right\| = \max_{a \le t \le b} \left| r(t) \right|. \tag{6}$$

Further, the ABC derivatives fulfil the Lipschitz condition [20]:

$$\left\| {}^{ABC}D_{0,t}^{\psi} r_1(t) - {}^{ABC}D_{0,t}^{\psi} r_2(t) \right\| \prec \omega \left\| r_1(t) - r_2(t) \right\|. \tag{7}$$

**2.2 Existence and Uniqueness**

We denote a Banach space by $D(W)$ with $W = [0,b]$, containing real valued continuous function with sup norm $T = D(W) \times D(W) \times D(W) \times D(W) \times D(W) \times D(W) \times D(W)$ and the given norm $\left\| (S, S_U, E, C, C_U, V, R) \right\| = \|S\| + \|S_U\| + \|E\| + \|C\| + \|C_U\| + \|V\| + \|R\|$, where $\|S\| = Sup_{t\in J}|S|, \|S_U\| = Sup_{t\in J}|S_U|, \|E\| = Sup_{t\in J}|E|, \|C\| = Sup_{t\in J}|C|, \|C_U\| = Sup_{t\in J}|C_U|, \|V\| = Sup_{t\in J}|V|,$ $\|R\| = Sup_{t\in J}|R|$. Using the $ABC$ integral operator on the system (1) we have

$$\begin{cases} S(t) - S(0) = {}^{ABC}D_{0,t}^{\psi}[S(t)]\{\Omega^{\psi} - \beta^{\psi}\left(\dfrac{C(t)+C_U(t)}{N(t)}\right)S - (\mu^{\psi} + \lambda^{\psi})S\}, \\ S_U(t) - S_U(0) = {}^{ABC}D_{0,t}^{\psi}[S_U(t)]\{\lambda^{\psi}S(t) + \gamma_1^{\psi}C(t) - \beta^{\psi}\left(\dfrac{C(t)+C_U(t)}{N(t)}\right)S_U(t) - (\delta_2^{\psi} + \mu^{\psi})S_U(t)\}, \\ E(t) - E(0) = {}^{ABC}D_{0,t}^{\psi}[\beta^{\psi}\left(\dfrac{C(t)+C_U(t)}{N(t)}\right)S + \beta^{\psi}\left(\dfrac{C(t)+C_U(t)}{N(t)}\right)S_U(t) - (\varphi^{\psi} + \mu^{\psi})E(t)\}, \\ C(t) - C(0) = {}^{ABC}D_{0,t}^{\psi}[\alpha\varphi^{\psi}E(t) - (\delta^{\psi} + \mu^{\psi} + \gamma^{\psi})C(t)\}, \\ C_U(t) - C(0) = {}^{ABC}D_{0,t}^{\psi}[(1-\alpha)\varphi^{\psi}E(t) - (\delta_1^{\psi} + \mu^{\psi})C_U(t)\}, \\ V(t) - V(0) = {}^{ABC}D_{0,t}^{\psi}[V(t)]\{\eta^{\psi}(S(t) + S_U(t)) - \mu^{\psi}V(t)\}, \\ R(t) - R(0) = {}^{ABC}D_{0,t}^{\psi}[R(t)]\{\gamma^{\psi}C(t) + (1-\gamma_1^{\psi})C_U(t) - \mu^{\psi}R(t)\}. \end{cases} \tag{8}$$

Now from Definition 1, we have

$$S(t)-S(0)=\frac{1-\psi}{B(\psi)}\Phi_1(\psi,t,S(t))+\frac{\psi}{B(\psi)\Gamma(\psi)}\times\int_0^t(t-\tau)^{\psi-1}\Phi_1(\psi,\tau,S(\tau))d\tau,$$

$$S_U(t)-S_U(0)=\frac{1-\psi}{B(\psi)}\Phi_2(\psi,t,U(t))+\frac{\psi}{B(\psi)\Gamma(\psi)}\times\int_0^t(t-\tau)^{\psi-1}\Phi_2(\psi,\tau,A(\tau))d\tau,$$

$$E(t)-E(0)=\frac{1-\psi}{B(\psi)}\Phi_3(\psi,t,E(t))+\frac{\psi}{B(\psi)\Gamma(\psi)}\times\int_0^t(t-\tau)^{\psi-1}\Phi_3(\psi,\tau,E(\tau))d\tau,$$

$$C(t)-C(0)=\frac{1-\psi}{B(\psi)}\Phi_4(\psi,t,C(t))+\frac{\psi}{B(\psi)\Gamma(\psi)}\times\int_0^t(t-\tau)^{\psi-1}\Phi_4(\psi,\tau,I_S(\tau))d\tau, \quad (9)$$

$$C_U(t)-C_U(0)=\frac{1-\psi}{B(\psi)}\Phi_5(\psi,t,C_U(t))+\frac{\psi}{B(\psi)\Gamma(\psi)}\times\int_0^t(t-\tau)^{\psi-1}\Phi_5(\psi,\tau,C_U(\tau))d\tau,$$

$$V(t)-V(0)=\frac{1-\psi}{B(\psi)}\Phi_6(\psi,t,V(t))+\frac{\psi}{B(\psi)\Gamma(\psi)}\times\int_0^t(t-\tau)^{\psi-1}\Phi_3(\psi,\tau,V(\tau))d\tau,$$

$$R(t)-R(0)=\frac{1-\psi}{B(\psi)}\Phi_7(\psi,t,R(t))+\frac{\psi}{B(\psi)\Gamma(\psi)}\times\int_0^t(t-\tau)^{\psi-1}\Phi_7(\psi,\tau,R(\tau))d\tau,$$

where

$$\Phi_1(\psi,\tau,S(t))=\Omega^{\psi}-\beta^{\psi}\left(\frac{C(t)+C_U(t)}{N(t)}\right)S(t)-(\mu^{\psi}+\lambda^{\psi})S(t),$$

$$\Phi_2(\psi,\tau,S_U(t))=\lambda^{\psi}S(t)+\gamma_1^{\psi}C(t)-\beta^{\psi}\left(\frac{C(t)+C_U(t)}{N(t)}\right)S_U(t)-(\delta_2^{\psi}+\mu^{\psi})S_U(t),$$

$$\Phi_3(\psi,\tau,E(t))=\beta^{\psi}\left(\frac{C(t)+C_U(t)}{N(t)}\right)S+\beta\left(\frac{C(t)+C_U(t)}{N(t)}\right)S_U(t)-(\varphi^{\psi}+\mu^{\psi})E(t),$$

$$\Phi_4(\psi,\tau,C(t))=\alpha\varphi^{\psi}E(t)-(\delta^{\psi}+\mu^{\psi}+\gamma^{\psi})C(t), \quad (10)$$

$$\Phi_5(\psi,\tau,C_U(t))=(1-\alpha)\varphi^{\psi}E(t)-(\delta_1^{\psi}+\mu^{\psi})C_U(t),$$

$$\Phi_6(\psi,\tau,V(t))=\eta^{\psi}(S(t)+S_U(t))-\mu^{\psi}V(t),$$

$$\Phi_7(\psi,\tau,R(t))=\gamma^{\psi}C+(1-\gamma_1^{\psi})C_U-\mu^{\psi}R(t).$$

Further, the Atangana-Baleanu-Caputo derivatives fulfil the Lipschitz condition [20] only if $S(t),S_U(t),E(t),C(t),C_U(t),V(t)$ and $R(t)$ possess an upper bound. Suppose $S(t)$ and $S^*(t)$ are couple functions, then

$$\left\|\Phi_1(\psi,t,S(t))-\Phi_1(\psi,t,S^*(t))\right\|=\left\|-\left[\beta^{\psi}\left(\frac{C(t)+C_U(t)}{N(t)}\right)S-(\mu^{\psi}+\lambda^{\psi})S\right](S(t)-S^*(t))\right\|. \quad (11)$$

Considering

$$d_1 = \left\| -\left( \beta^{\psi} \left( \frac{C(t)+C_U(t)}{N(t)} \right) S - (\mu^{\psi} + \lambda^{\psi}) S \right) \right\|,$$

Equation (11) simplifies to

$$\left\| \Phi_1(\psi,t,S(t)) - \Phi_1(\psi,t,S^*(t)) \right\| \le d_1 \left\| (S(t) - S^*(t)) \right\|. \qquad (12)$$

Similarly,

$$\begin{aligned}
&\left\| \Phi_2(\psi,t,S_U(t)) - \Phi_2(\psi,t,S_U^*(t)) \right\| \le d_2 \left\| (S_U(t) - S_U^*(t)) \right\|,\\
&\left\| \Phi_3(\psi,t,E(t)) - \Phi_3(\psi,t,E^*(t)) \right\| \le d_3 \left\| (E(t) - E^*(t)) \right\|,\\
&\left\| \Phi_4(\psi,t,C(t)) - \Phi_4(\psi,t,C^*(t)) \right\| \le d_4 \left\| (C(t) - C^*(t)) \right\|,\\
&\left\| \Phi_5(\psi,t,C_U(t)) - \Phi_5(\psi,t,C_U^*(t)) \right\| \le d_5 \left\| (C_U(t) - C_U^*(t)) \right\|,\\
&\left\| \Phi_6(\psi,t,V(t)) - \Phi_6(\psi,t,V^*(t)) \right\| \le d_6 \left\| (V(t) - V^*(t)) \right\|,\\
&\left\| \Phi_7(\psi,t,R(t)) - \Phi_7(\psi,t,R^*(t)) \right\| \le d_7 \left\| (R(t) - R^*(t)) \right\|.
\end{aligned} \qquad (13)$$

Where

$$\begin{aligned}
d_2 &= \left\| -\beta^{\psi} \left( \frac{C(t)+C_U(t)}{N(t)} \right) S_U(t) - (\delta_2^{\psi} + \mu^{\psi}) S_U(t) \right\|,\\
d_3 &= \left\| -(\varphi^{\psi} + \mu^{\psi}) E(t) \right\|,\\
d_4 &= \left\| (\delta^{\psi} + \mu^{\psi} + \gamma^{\psi}) C(t) \right\|,\\
d_5 &= \left\| -(\delta_1^{\psi} + \mu^{\psi}) C_U(t) \right\|,\\
d_6 &= \left\| -\mu^{\psi} \right\|,\\
d_7 &= \left\| -\mu^{\psi} \right\|.
\end{aligned}$$

Hence Lipschitz condition holds. Now taking system (9) in a reiterative manner gives

$$S_n(t)-S(0)=\frac{1-\psi}{B(\psi)}\Phi_1(\psi,t,S_{n-1}(t))+\frac{\psi}{B(\psi)\Gamma(\psi)}\times\int_0^t(t-\tau)^{\psi-1}\Phi_1(\psi,\tau,S_{n-1}(\tau))d\tau,$$

$$S_{U\,n}(t)-S_U(0)=\frac{1-\psi}{B(\psi)}\Phi_2(\psi,t,E_{n-1}(t))+\frac{\psi}{B(\psi)\Gamma(\psi)}\times\int_0^t(t-\tau)^{\psi-1}\Phi_2(\psi,\tau,E_{n-1}(\tau))d\tau,$$

$$E(t)-E(0)=\frac{1-\psi}{B(\psi)}\Phi_3(\psi,t,I_{A_{n-1}}(t))+\frac{\psi}{B(\psi)\Gamma(\psi)}\times\int_0^t(t-\tau)^{\psi-1}\Phi_3(\psi,\tau,I_{A_{n-1}}(\tau))d\tau,$$

$$C(t)-C(0)=\frac{1-\psi}{B(\psi)}\Phi_4(\psi,t,I_{S_{n-1}}(t))+\frac{\psi}{B(\psi)\Gamma(\psi)}\times\int_0^t(t-\tau)^{\psi-1}\Phi_4(\psi,\tau,I_{S_{n-1}}(\tau))d\tau, \tag{14}$$

$$C_{U\,n}(t)-C_U(0)=\frac{1-\psi}{B(\psi)}\Phi_5(\psi,t,Q_{n-1}(t))+\frac{\psi}{B(\psi)\Gamma(\psi)}\times\int_0^t(t-\tau)^{\psi-1}\Phi_5(\psi,\tau,Q_{n-1}(\tau))d\tau,$$

$$V_n(t)-V(0)=\frac{1-\psi}{B(\psi)}\Phi_6(\psi,t,V_{n-1}(t))+\frac{\psi}{B(\psi)\Gamma(\psi)}\times\int_0^t(t-\tau)^{\psi-1}\Phi_5(\psi,\tau,V_{n-1}(\tau))d\tau,$$

$$R_n(t)-R(0)=\frac{1-\psi}{B(\psi)}\Phi_7(\psi,t,R_{n-1}(t))+\frac{\psi}{B(\psi)\Gamma(\psi)}\times\int_0^t(t-\tau)^{\psi-1}\Phi_7(\psi,\tau,R_{n-1}(\tau))d\tau,$$

with the initial conditions

$$S(0)=S_0, S_U(0)=S_{U\,0}, E(0)=E_0, C(0)=C_0, C_U(0)=C_{U\,0}, V(0)=V_0, R(0)=R_0.$$

Difference of consecutive terms yields

$$\Pi_{S_n}(t) = S_n(t) - S_{n-1}(t) = \frac{1-\psi}{B(\psi)}(\Phi_1(\psi,t,S_{n-1}(t)) - \Phi_1(\psi,t,S_{n-2}(t)))$$

$$+\frac{\psi}{B(\psi)\Gamma(\psi)}\int_0^t (t-\tau)^{\psi-1}(\Phi_1(\psi,\tau,S_{n-1}(\tau)) - \Phi_1(\psi,\tau,S_{n-2}(\tau)))d\tau,$$

$$\Pi_{S_{U_n}}(t) = S_{U\,n}(t) - S_{U\,n-1}(t) = \frac{1-\psi}{B(\psi)}(\Phi_2(\psi,t,S_{U\,n-1}(t)) - \Phi_2(\psi,t,S_{U\,n-2}(t)))$$

$$+\frac{\psi}{B(\psi)\Gamma(\psi)}\int_0^t (t-\tau)^{\psi-1}(\Phi_2(\psi,\tau,S_{U\,n-1}(\tau)) - \Phi_2(\psi,\tau,S_{U\,n-2}(\tau)))d\tau,$$

$$\Pi_{E_n}(t) = E_n(t) - E_{n-1}(t) = \frac{1-\psi}{B(\psi)}(\Phi_3(\psi,t,E_{n-1}(t)) - \Phi_3(\psi,t,E_{n-2}(t)))$$

$$+\frac{\psi}{B(\psi)\Gamma(\psi)}\int_0^t (t-\tau)^{\psi-1}(\Phi_3(\psi,\tau,E_{n-1}(\tau)) - \Phi_3(\psi,\tau,E_{n-2}(\tau)))d\tau,$$

$$\Pi_{C_n}(t) = C_n(t) - C_{n-1}(t) = \frac{1-\psi}{B(\psi)}(\Phi_4(\psi,t,C_{n-1}(t)) - \Phi_4(\psi,t,C_{n-2}(t)))$$

$$+\frac{\psi}{B(\psi)\Gamma(\psi)}\int_0^t (t-\tau)^{\psi-1}(\Phi_4(\psi,\tau,C_{n-1}(\tau)) - \Phi_4(\psi,\tau,C_{n-2}(\tau)))d\tau,$$

$$\Pi_{C_{U_n}}(t) = C_{U\,n}(t) - C_{U\,n-1}(t) = \frac{1-\psi}{B(\psi)}(\Phi_5(\psi,t,C_{U\,n-1}(t)) - \Phi_5(\psi,t,C_{U\,n-2}(t)))$$

$$+\frac{\psi}{B(\psi)\Gamma(\psi)}\int_0^t (t-\tau)^{\psi-1}(\Phi_5(\psi,\tau,C_{U\,n-1}(\tau)) - \Phi_5(\psi,\tau,C_{U\,n-2}(\tau)))d\tau,$$

$$\Pi_{V_n}(t) = V_n(t) - V_{n-1}(t) = \frac{1-\psi}{B(\psi)}(\Phi_6(\psi,t,V_{n-1}(t)) - \Phi_6(\psi,t,V_{n-2}(t))) \tag{15}$$

$$+\frac{\psi}{B(\psi)\Gamma(\psi)}\int_0^t (t-\tau)^{\psi-1}(\Phi_6(\psi,\tau,V_{n-1}(\tau)) - \Phi_6(\psi,\tau,V_{n-2}(\tau)))d\tau,$$

$$\Pi_{R_n}(t) = R_n(t) - R_{n-1}(t) = \frac{1-\psi}{B(\psi)}(\Phi_7(\psi,t,R_{n-1}(t)) - \Phi_7(\psi,t,R_{n-2}(t)))$$

$$+\frac{\psi}{B(\psi)\Gamma(\psi)}\int_0^t (t-\tau)^{\psi-1}(\Phi_7(\psi,\tau,R_{n-1}(\tau)) - \Phi_7(\psi,\tau,R_{n-2}(\tau)))d\tau,$$

where $S_n(t) = \sum_{i=0}^{n}\Pi_{S_n}(t), S_{U\,n}(t) = \sum_{i=0}^{n}\Pi_{S_{Un}}(t), E_n(t) = \sum_{i=0}^{n}\Pi_{E_n}(t), C(t) = \sum_{i=0}^{n}\Pi_C(t),$ $C_{U\,n}(t) = \sum_{i=0}^{n}\Pi_{C_{U_n}}(t), V_n(t) = \sum_{i=0}^{n}\Pi_{V_n}(t), R_n(t) = \sum_{i=0}^{n}\Pi_{R_n}(t)$. Taking into consideration equation (12) – (13) and considering $\Pi_{S_{n-1}}(t) = S_{n-1}(t) - S_{n-2}(t), \Pi_{E_{n-1}}(t) = E_{n-1}(t) - E_{n-2}(t),$

$\Pi_C(t)=C_{n-1}(t)-C_{n-2}(t),\Pi_{C_{U\,n-1}}(t)=C_{U\,n-1}(t)-C_{U\,n-2}(t),\Pi_{V_{n-1}}(t)=V_{n-1}(t)-V_{n-2}(t),$
$\Pi_{R_{n-1}}(t)=R_{n-1}(t)-R_{n-2}(t),$

$$
\begin{aligned}
&\left\|\Pi_{S_n}(t)\right\|\le\frac{1-\psi}{B(\psi)}d_1\left\|\Pi_{S_{n-1}}(t)\right\|\frac{\psi}{B(\psi)\Gamma(\psi)}d_1\times\int_0^t(t-\tau)^{\psi-1}\left\|\Pi_{S_{n-1}}(\tau)\right\|d\tau,\\
&\left\|\Pi_{S_{U\,n}}(t)\right\|\le\frac{1-\psi}{B(\psi)}d_2\left\|\Pi_{S_{U\,n-1}}(t)\right\|\frac{\psi}{B(\psi)\Gamma(\psi)}d_2\times\int_0^t(t-\tau)^{\psi-1}\left\|\Pi_{S_{U\,n-1}}(\tau)\right\|d\tau,\\
&\left\|\Pi_{E_n}(t)\right\|\le\frac{1-\psi}{B(\psi)}d_3\left\|\Pi_{E_{n-1}}(t)\right\|\frac{\psi}{B(\psi)\Gamma(\psi)}d_3\times\int_0^t(t-\tau)^{\psi-1}\left\|\Pi_{E_{n-1}}(\tau)\right\|d\tau,\\
&\left\|\Pi_{C_n}(t)\right\|\le\frac{1-\psi}{B(\psi)}d_4\left\|\Pi_{C_{n-1}}(t)\right\|\frac{\psi}{B(\psi)\Gamma(\psi)}d_4\times\int_0^t(t-\tau)^{\psi-1}\left\|\Pi_{C_{n-1}}(\tau)\right\|d\tau,\\
&\left\|\Pi_{C_{U\,n}}(t)\right\|\le\frac{1-\psi}{B(\psi)}d_5\left\|\Pi_{C_{U\,n-1}}(t)\right\|\frac{\psi}{B(\psi)\Gamma(\psi)}d_5\times\int_0^t(t-\tau)^{\psi-1}\left\|\Pi_{C_{U\,n-1}}(\tau)\right\|d\tau,\\
&\left\|\Pi_{V_n}(t)\right\|\le\frac{1-\psi}{B(\psi)}d_6\left\|\Pi_{V_{n-1}}(t)\right\|\frac{\psi}{B(\psi)\Gamma(\psi)}d_6\times\int_0^t(t-\tau)^{\psi-1}\left\|\Pi_{V_{n-1}}(\tau)\right\|d\tau,\\
&\left\|\Pi_{R_n}(t)\right\|\le\frac{1-\psi}{B(\psi)}d_7\left\|\Pi_{R_{n-1}}(t)\right\|\frac{\psi}{B(\psi)\Gamma(\psi)}d_7\times\int_0^t(t-\tau)^{\psi-1}\left\|\Pi_{R_{n-1}}(\tau)\right\|d\tau.
\end{aligned}
\tag{16}
$$

**Theorem 2:** The system (1) has a unique solution for $t\in[0,b]$ subject to the condition $\frac{1-\psi}{B(\psi)}d_i+\frac{\psi}{B(\psi)\Gamma(\psi)}b^{\psi}n_i\prec 1, i=1,2,3,.......,7$ hold.

**Proof:**

Since $S(t),S_U(t),E(t),C(t),C_U(t),V(t)$ and $R(t)$ are bounded functions and Equation $(12)-(13)$ holds. In a recurring manner (16) reaches

$$\left\|\Pi_{S_n}(t)\right\| \le \left\|S_o(t)\right\|\left(\frac{1-\psi}{B(\psi)}d_1 + \frac{\psi b^{\psi}}{B(\psi)\Gamma(\psi)}d_1\right)^n,$$

$$\left\|\Pi_{S_{U_n}}(t)\right\| \le \left\|S_{U\,o}(t)\right\|\left(\frac{1-\psi}{B(\psi)}d_2 + \frac{\psi b^{\psi}}{B(\psi)\Gamma(\psi)}d_2\right)^n,$$

$$\left\|\Pi_{E_n}(t)\right\| \le \left\|E_o(t)\right\|\left(\frac{1-\psi}{B(\psi)}d_3 + \frac{\psi b^{\psi}}{B(\psi)\Gamma(\psi)}d_3\right)^n, \qquad (17)$$

$$\left\|\Pi_{C_{S_n}}(t)\right\| \le \left\|C_o(t)\right\|\left(\frac{1-\psi}{B(\psi)}d_4 + \frac{\psi b^{\psi}}{B(\psi)\Gamma(\psi)}d_4\right)^n,$$

$$\left\|\Pi_{C_{U_n}}(t)\right\| \le \left\|C_{U\,o}(t)\right\|\left(\frac{1-\psi}{B(\psi)}d_5 + \frac{\psi b^{\psi}}{B(\psi)\Gamma(\psi)}d_5\right)^n,$$

$$\left\|\Pi_{V_n}(t)\right\| \le \left\|V_o(t)\right\|\left(\frac{1-\psi}{B(\psi)}d_6 + \frac{\psi b^{\psi}}{B(\psi)\Gamma(\psi)}d_6\right)^n,$$

$$\left\|\Pi_{R_n}(t)\right\| \le \left\|R_o(t)\right\|\left(\frac{1-\psi}{B(\psi)}d_7 + \frac{\psi b^{\psi}}{B(\psi)\Gamma(\psi)}d_7\right)^n,$$

and

$$\left\|\Pi_{S_n}(t)\right\| \to 0, \left\|\Pi_{S_{U_n}}(t)\right\| \to 0, \left\|\Pi_{E_n}(t)\right\| \to 0, \left\|\Pi_{C_n}(t)\right\| \to 0, \left\|\Pi_{C_{U_n}}(t)\right\| \to 0, \left\|\Pi_{V_n}(t)\right\| \to 0, \left\|\Pi_{R_n}(t)\right\| \to 0$$

as $n \to \infty$. Incorporating the triangular inequality and for any $j$, system (17) yields

$$\left\|S_{n+j}(t) - S_n(t)\right\| \le \sum_{i=n+1}^{n+j} T_1^{\,j} = \frac{T_1^{n+1} - T_1^{n+k+1}}{1-T_1},$$

$$\left\|S_{U\,n+j}(t) - S_{U\,n}(t)\right\| \le \sum_{i=n+1}^{n+j} T_2^{\,j} = \frac{T_2^{n+1} - T_2^{n+k+1}}{1-T_2},$$

$$\left\|E_{n+j}(t) - E_n(t)\right\| \le \sum_{i=n+1}^{n+j} T_3^{\,j} = \frac{T_3^{n+1} - T_3^{n+k+1}}{1-T_3},$$

$$\left\|C_{n+j}(t) - C_n(t)\right\| \le \sum_{i=n+1}^{n+j} T_4^{\,j} = \frac{T_4^{n+1} - T_4^{n+k+1}}{1-T_4}, \qquad (18)$$

$$\left\|C_{U\,n+j}(t) - C_{U\,n}(t)\right\| \le \sum_{i=n+1}^{n+j} T_5^{\,j} = \frac{T_5^{n+1} - T_5^{n+k+1}}{1-T_5},$$

$$\left\|V_{n+j}(t) - V_n(t)\right\| \le \sum_{i=n+1}^{n+j} T_6^{\,j} = \frac{T_6^{n+1} - T_6^{n+k+1}}{1-T_6},$$

$$\left\|R_{n+j}(t) - R_n(t)\right\| \le \sum_{i=n+1}^{n+j} T_7^{\,j} = \frac{T_7^{n+1} - T_7^{n+k+1}}{1-T_7}.$$

Where $T_i = \frac{1-\psi}{B(\psi)} d_i + \frac{\psi}{B(\psi)\Gamma(\psi)} b^{\psi} d_i \prec 1$. Hence there exists a unique solution for system (1)

## 3 Analysis of the Model

### 3.1 The Steady States of the Model

The disease-free equilibrium is the steady-state solution where there is no COVID-19 infection in the population. Setting $E = I = C = 0$ and the right hand side of the system (1) to zero, then solving yields

$$E^0 = (S^0, S_U{}^0, E^0, C^0, C_U{}^0, V^0, R^0) = \left( \frac{\Omega^{\psi}}{\mu^{\psi} + \lambda^{\psi}}, \frac{\lambda^{\psi}}{(\delta_2^{\psi} + \mu^{\psi})}, 0, 0, 0, \frac{\eta^{\psi}(S + S_U)}{\mu^{\psi}}, 0 \right). \qquad (19)$$

The endemic equilibrium point, $\mathrm{E}_* = (S^*, D^*, E^*, I^*, C^*, R^*)$ is given as

$$\begin{cases} S^* = \dfrac{\Omega^{\psi}}{\beta^{\psi} \dfrac{(C^* + C_U{}^*)}{N} + (\mu^{\psi} + \lambda^{\psi})}, S_U{}^* = \dfrac{\lambda^{\psi} S^* + \gamma_1{}^{\psi} C^*}{\beta^{\psi} \dfrac{(I^* + C^*)}{N} + (\mu^{\psi} + \delta_2{}^{\psi})}, \\ E^* = \dfrac{\beta^{\psi}}{(\varphi^{\psi} + \mu^{\psi})} \left[ \dfrac{(C^* + C_U{}^*)(S^* + S_U{}^*)}{N} \right], C^* = \dfrac{\alpha \varphi^{\psi} E^*}{(\delta^{\psi} + \mu^{\psi} + \gamma^{\psi})}, \\ C_U{}^* = \dfrac{(1-\alpha)\varphi^{\psi} E^*}{(\delta_1{}^{\psi} + \mu^{\psi} + \gamma_1{}^{\psi})}, V^* = \dfrac{\eta^{\psi}(S^* + S_U{}^*)}{\mu^{\psi}}, R^* = \dfrac{\gamma I^*}{\mu}. \end{cases} \qquad (20)$$

### 4.1 Basic Reproductive Number

We now calculate the basic reproductive number $(R_o)$ of system (1). The basic reproductive number is the number of secondary cases produced, in a susceptible population, by a single infective individual during the time of the infection. We evaluate the basic reproductive number using the next generation operator method [21]. From system (1), $E, C$ and $C_U$ are the COVID-19 infected compartments. We decomposed the right-hand side of system (1) corresponding to the COVID-19 infected compartments as $F - Z$, where

$$F = \begin{pmatrix} \beta^{\psi}\left(\frac{C + C_U}{N}\right)S + \beta^{\psi}\left(\frac{C + C_U}{N}\right)D \\ \alpha\varphi^{\psi} E \\ (1-\alpha)\varphi^{\psi} E \end{pmatrix} \text{ and } Z = \begin{pmatrix} (\varphi^{\psi} + \mu^{\psi})E \\ (\delta^{\psi} + \mu^{\psi} + \gamma^{\psi})C \\ (\delta_1^{\psi} + \mu^{\psi} + \gamma_1^{\psi})C_U \end{pmatrix}.$$

Next, we find the derivative of $F$ and $Z$ evaluated at the disease-free steady state and this gives the matrices

$$F = \frac{\partial F}{\partial x_i} = \begin{pmatrix} 0 & \frac{\beta^{\psi}(S^0 + D^0)}{N} & \frac{\beta^{\psi}(S^0 + D^0)}{N} \\ \alpha\varphi^{\psi} & 0 & 0 \\ (1-\alpha)\varphi^{\psi} & 0 & 0 \end{pmatrix} \text{and}$$

$$Z = \frac{\partial Z}{\partial x_i} = \begin{pmatrix} \varphi^{\psi} + \mu^{\psi} & 0 & 0 \\ 0 & \delta^{\psi} + \mu^{\psi} + \gamma^{\psi} & 0 \\ 0 & 0 & \delta_1^{\psi} + \mu^{\psi} + \gamma_1^{\psi} \end{pmatrix}.$$

Where $x_i = E, C, C_U$

$$FZ^{-1} = \begin{pmatrix} 0 & (\delta^{\psi} + \mu^{\psi} + \gamma^{\psi})\beta^{\psi}\left(\frac{S^0 + D^0}{N}\right) & 0 \\ \alpha\varphi^{\psi}(\varphi^{\psi} + \mu^{\psi}) & 0 & 0 \\ (1-\alpha)\varphi^{\psi} & 0 & 0 \end{pmatrix}. \qquad (21)$$

The basic reproductive number is the largest positive eigenvalue of $FZ^{-1}$ and is given as

$$R_0 = \sqrt{\beta^{\psi}\alpha\varphi^{\psi}(\varphi^{\psi}+\mu^{\psi})\left(\delta^{\psi}+\mu^{\psi}+\gamma^{\psi}\right)\left(\frac{\Omega^{\psi}(\lambda^{\psi}+\delta_2{}^{\psi}+\mu^{\psi})}{(\mu^{\psi}+\lambda^{\psi})(\delta_2{}^{\psi}+\mu^{\psi})}\right)+(1-\alpha)\varphi^{\psi}}\ . \qquad (22)$$

## 4.2 Local Stability of the Equilibrium Points

The necessary condition for the local stability of both the disease –free and endemic steady state is established in Theorem 3.

**Theorem 3:** The disease-free equilibrium, if it exists, is locally asymptotically stable if $R_o < 1$ and unstable when $R_o > 1$ .

**Proof:**

The Jacobian matrix of system (1) evaluated at the disease-free equilibrium point is

$$J = \begin{pmatrix} J_{11} & 0 & 0 & J_{14} & J_{15} & 0 & 0 \\ \lambda^{\psi} & J_{22} & 0 & J_{24} & J_{25} & 0 & 0 \\ 0 & 0 & -(\varphi^{\psi}+\mu^{\psi}) & J_{34} & J_{35} & 0 & 0 \\ 0 & 0 & \alpha\varphi^{\psi} & J_{44} & 0 & 0 & 0 \\ 0 & 0 & J_{53} & 0 & J_{55} & 0 & 0 \\ \eta^{\psi} & \eta^{\psi} & 0 & 0 & 0 & -\mu^{\psi} & 0 \\ 0 & 0 & 0 & \gamma^{\psi} & 0 & 0 & -\mu^{\psi} \end{pmatrix}. \qquad (23)$$

Where

$$J_{11} = -(\lambda^{\psi}+\mu^{\psi}), J_{22} = -(\delta_2{}^{\psi}+\mu^{\psi}), J_{53} = (1-\alpha)\varphi^{\psi}, J_{34} = J_{35} = \frac{\beta^{\psi}(S^o+S_U^o)}{N}, J_{25} = \gamma_1^{\psi} - \frac{\beta^{\psi}S_U^o}{N}$$

$$J_{14} = J_{15} = -\beta^{\psi}\frac{S^0}{N}, J_{24} = -\frac{\beta^{\psi}S_U^o}{N}, J_{55} = -(\delta^{\psi}+\mu^{\psi}+\gamma_1^{\psi}), J_{44} = -(\delta^{\psi}+\gamma^{\psi}+\mu^{\psi}).$$

The first four (4) eigenvalues are $-(\lambda^{\psi}+\mu^{\psi}), -(\delta_2{}^{\psi}+\mu^{\psi})$ and $-\mu$. The remaining ones can be obtained by deleting the first, second, sixth and seventh rows and columns of the system (23).

Hence we have

$$J_{E^0} = \begin{bmatrix} -\varphi^{\psi} - \mu^{\psi} & J_{34} & J_{35} \\ \alpha\varphi^{\psi} & J_{44} & 0 \\ J_{53} & 0 & -(\delta_1 + \mu) \end{bmatrix}. \quad (24)$$

The characteristic equation of system (24) is

$$\Phi(\omega) = \omega^3 + A_1\omega^2 + A_2\omega + A_3 = 0. \quad (25)$$

Where

$$A_1 = -(\delta + \mu + \gamma)(\delta_1 + \mu)(\varphi + \mu),$$
$$A_2 = (\varphi + \mu)[(\delta + \mu + \gamma) + (\delta_1 + \mu) - A\varphi + (\delta + \mu + \gamma)(\delta_1 + \mu)],$$
$$A_3 = -\beta^{\psi} \frac{S^0}{N}((\delta + \mu + \gamma)(1-\alpha)\varphi - \alpha\varphi(\delta_1 + \mu)) + (\delta + \mu + \gamma)(\delta_1 + \mu)(\varphi + \mu).$$

Let $R_o > 0$, then $A_2 < 0$ and $A_3 < 0$. This shows that the system (1) has a unique endemic steady state which is stable. In addition from Routh – Hurwitz stability criterion, if the conditions $A_1 > 0, A_3 > 0$ and $A_1A_2 - A_3 > 0$ are true, then all the roots of the characteristic equation (25) have a negative real part which means stable equilibrium but clearly, $A_1 < 0$ and this makes the disease-free steady state unstable since we already having a group with an underlying condition in the population.

## 5 Hyers –Ulam Stability

**Definition 4**

The ABC fractional system given by equation (1) is said to be Hyers-Ulam stable if, for every $\lambdabar_i \succ 0, i \in N^7$, there exist constants $\hbar_i \succ 0, i \in N^7$ satisfying:

$$\left| S(t) - \frac{1-\psi}{B(\psi)}\Phi_1(\psi,t,S(t)) + \frac{\psi}{B(\psi)\Gamma(\psi)} \times \int_0^t (t-\tau)^{\psi-1} \Phi_1(\psi,\tau,S(\tau))d\tau \right| \leq \lambda_1,$$

$$\left| S_U(t) - \frac{1-\psi}{B(\psi)}\Phi_2(\psi,t,S_U(t)) + \frac{\psi}{B(\psi)\Gamma(\psi)} \times \int_0^t (t-\tau)^{\psi-1} \Phi_2(\psi,\tau,S_U(\tau))d\tau \right| \leq \lambda_2,$$

$$\left| E(t) - \frac{1-\psi}{B(\psi)}\Phi_3(\psi,t,E(t)) + \frac{\psi}{B(\psi)\Gamma(\psi)} \times \int_0^t (t-\tau)^{\psi-1} \Phi_3(\psi,\tau,E(\tau))d\tau \right| \leq \lambda_3,$$

$$\left| C(t) - \frac{1-\psi}{B(\psi)}\Phi_4(\psi,t,C(t)) + \frac{\psi}{B(\psi)\Gamma(\psi)} \times \int_0^t (t-\tau)^{\psi-1} \Phi_4(\psi,\tau,C(\tau))d\tau \right| \leq \lambda_4, \qquad (26)$$

$$\left| C_U(t) - \frac{1-\psi}{B(\psi)}\Phi_5(\psi,t,C_U(t)) + \frac{\psi}{B(\psi)\Gamma(\psi)} \times \int_0^t (t-\tau)^{\psi-1} \Phi_5(\psi,\tau,C_U(\tau))d\tau \right| \leq \lambda_5,$$

$$\left| V(t) - \frac{1-\psi}{B(\psi)}\Phi_6(\psi,t,V(t)) + \frac{\psi}{B(\psi)\Gamma(\psi)} \times \int_0^t (t-\tau)^{\psi-1} \Phi_6(\psi,\tau,V(\tau))d\tau \right| \leq \lambda_6,$$

$$\left| R(t) - \frac{1-\psi}{B(\psi)}\Phi_7(\psi,t,R(t)) + \frac{\psi}{B(\psi)\Gamma(\psi)} \times \int_0^t (t-\tau)^{\psi-1} \Phi_7(\psi,\tau,R(\tau))d\tau \right| \leq \lambda_7,$$

And there exist $\{\dot{S}(t), \dot{S}_U(t), \dot{E}(t), \dot{C}(t), \dot{C}_U(t), \dot{V}(t), \dot{R}(t)\}$ where

$$\dot{S}(t) = \frac{1-\psi}{B(\psi)}\Phi_1(\psi,t,S(t)) + \frac{\psi}{B(\psi)\Gamma(\psi)} \times \int_0^t (t-\tau)^{\psi-1} \Phi_1(\psi,\tau,\dot{S}(\tau))d\tau,$$

$$\dot{S}_U(t) = \frac{1-\psi}{B(\psi)}\Phi_2(\psi,t,S_U(t)) + \frac{\psi}{B(\psi)\Gamma(\psi)} \times \int_0^t (t-\tau)^{\psi-1} \Phi_2(\psi,\tau,\dot{S}_U(\tau))d\tau,$$

$$\dot{E}(t) = \frac{1-\psi}{B(\psi)}\Phi_3(\psi,t,E(t)) + \frac{\psi}{B(\psi)\Gamma(\psi)} \times \int_0^t (t-\tau)^{\psi-1} \Phi_3(\psi,\tau,\dot{E}(\tau))d\tau,$$

$$\dot{C}(t) = \frac{1-\psi}{B(\psi)}\Phi_4(\psi,t,C(t)) + \frac{\psi}{B(\psi)\Gamma(\psi)} \times \int_0^t (t-\tau)^{\psi-1} \Phi_4(\psi,\tau,\dot{C}(\tau))d\tau, \qquad (27)$$

$$\dot{C}_U(t) = \frac{1-\psi}{B(\psi)}\Phi_5(\psi,t,C_U(t)) + \frac{\psi}{B(\psi)\Gamma(\psi)} \times \int_0^t (t-\tau)^{\psi-1} \Phi_3(\psi,\tau,\dot{C}_U(\tau))d\tau,$$

$$\dot{V}(t) = \frac{1-\psi}{B(\psi)}\Phi_6(\psi,t,V(t)) + \frac{\psi}{B(\psi)\Gamma(\psi)} \times \int_0^t (t-\tau)^{\psi-1} \Phi_6(\psi,\tau,\dot{V}(\tau))d\tau,$$

$$\dot{R}(t) = \frac{1-\psi}{B(\psi)}\Phi_7(\psi,t,R(t)) + \frac{\psi}{B(\psi)\Gamma(\psi)} \times \int_0^t (t-\tau)^{\psi-1} \Phi_7(\psi,\tau,\dot{R}(\tau))d\tau.$$

Such that

$$\left|S(t)-\overset{\bullet}{S}(t)\right| \le \xi_1 \lambda_1, \left|S_U(t)-\overset{\bullet}{S}_U(t)\right| \le \xi_2 \lambda_2, \left|E(t)-\overset{\bullet}{E}(t)\right| \le \xi_3 \lambda_3, \left|C(t)-\overset{\bullet}{C}(t)\right| \le \xi_4 \lambda_4,$$

$$\left|C_U(t)-\overset{\bullet}{C}_U(t)\right| \le \xi_5 \lambda_5, \left|V(t)-\overset{\bullet}{V}(t)\right| \le \xi_6 \lambda_6, \left|R(t)-\overset{\bullet}{R}(t)\right| \le \xi_7 \lambda_7.$$

## 6 Numerical Simulation

In this section, we validate the model using COVID-19 cases data of Ghana between March and June 2020 [24]. The COVID-19 cumulative cases with the best-fitted curve are given in figure 1. The parameter values which are estimated and some taken from the literature are given in table 1.

Table 1: Parameter values and description

| Parameter | Description | Value | Source |
|---|---|---|---|
| $\Omega^{\psi}$ | Recruitment rate | 29.08 | [11] |
| $\beta^{\psi}$ | COVID-19 transmission rate | 0.9 | Estimated |
| $\varphi^{\psi}$ | The rate at which exposed individuals become infectious | 0.0000230757 | [32] |
| $\mu^{\psi}$ | Natural death rate | $0.4252912\times10^{-4}$ | Estimated |
| $\delta^{\psi}$ | COVID-19 disease-induced death rate | $1.6728\times10^{-5}$ | [12] |
| $\gamma_1^{\psi}$ | The recovery rate of COVID-19 patients with an underlying condition | $1/14$ | Estimated |
| $\delta_2^{\psi}$ | Disease-induced death rate of an underlying condition | 0.05 | [39] |
| $\lambda^{\psi}$ | Rate of getting the underlying condition | 0.2 | [10] |
| $\gamma^{\psi}$ | The recovery rate of COVID-19 patients without an underlying condition | 1/14 | Estimated |
| $\delta_1^{\psi}$ | The disease-induced death rate of COVID-19 with an underlying condition | 0.0144 | [40, 41] |
| $\eta^{\psi}$ | The rate at which susceptible individuals are vaccinated | 0.0269 | [13] |

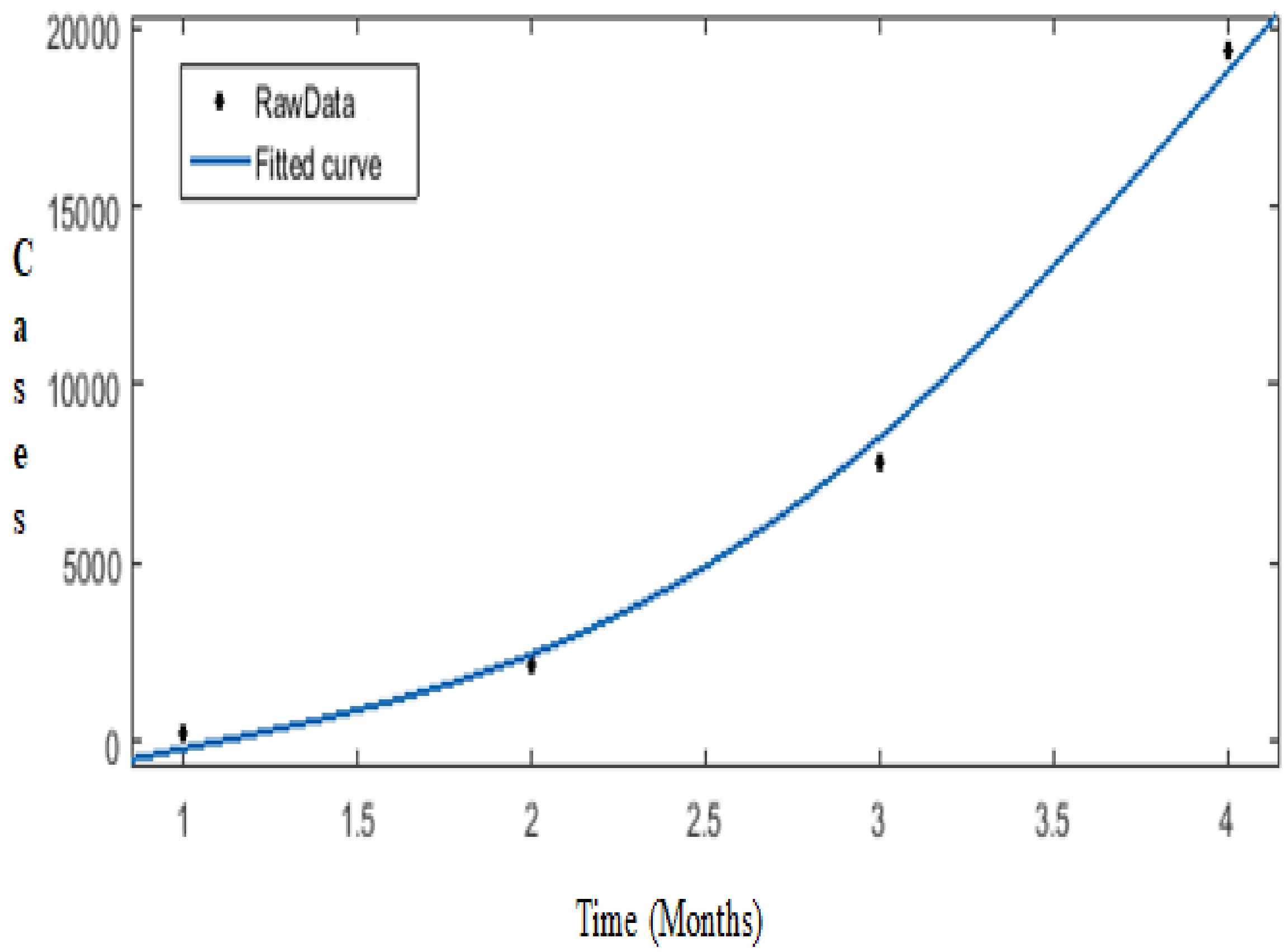

Fig.2: Cumulative cases of Ghana's COVID-19 from March - September 2020 with the best-fitted curve.

Using the initial conditions $S(0)=100000, S_U(0)=100, E(0)=10, C(0)=10, C_U(0)=10, V(0)=0, R(0)=0$ and the parameter values given in Table 1, the results of the simulation are displayed in Figs. 3 – 8.

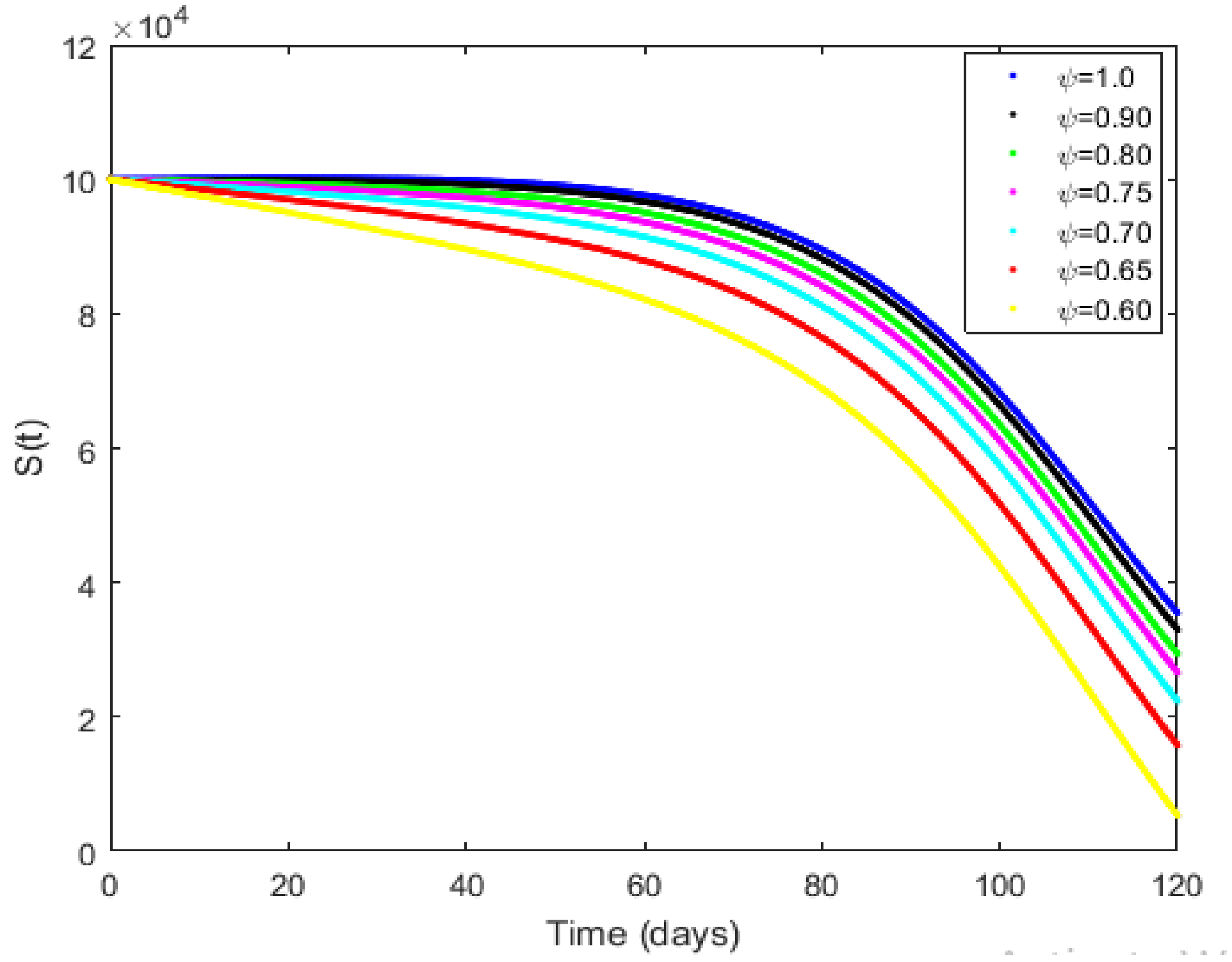

Fig.3: Behaviour of the susceptible population given different values of $\psi$

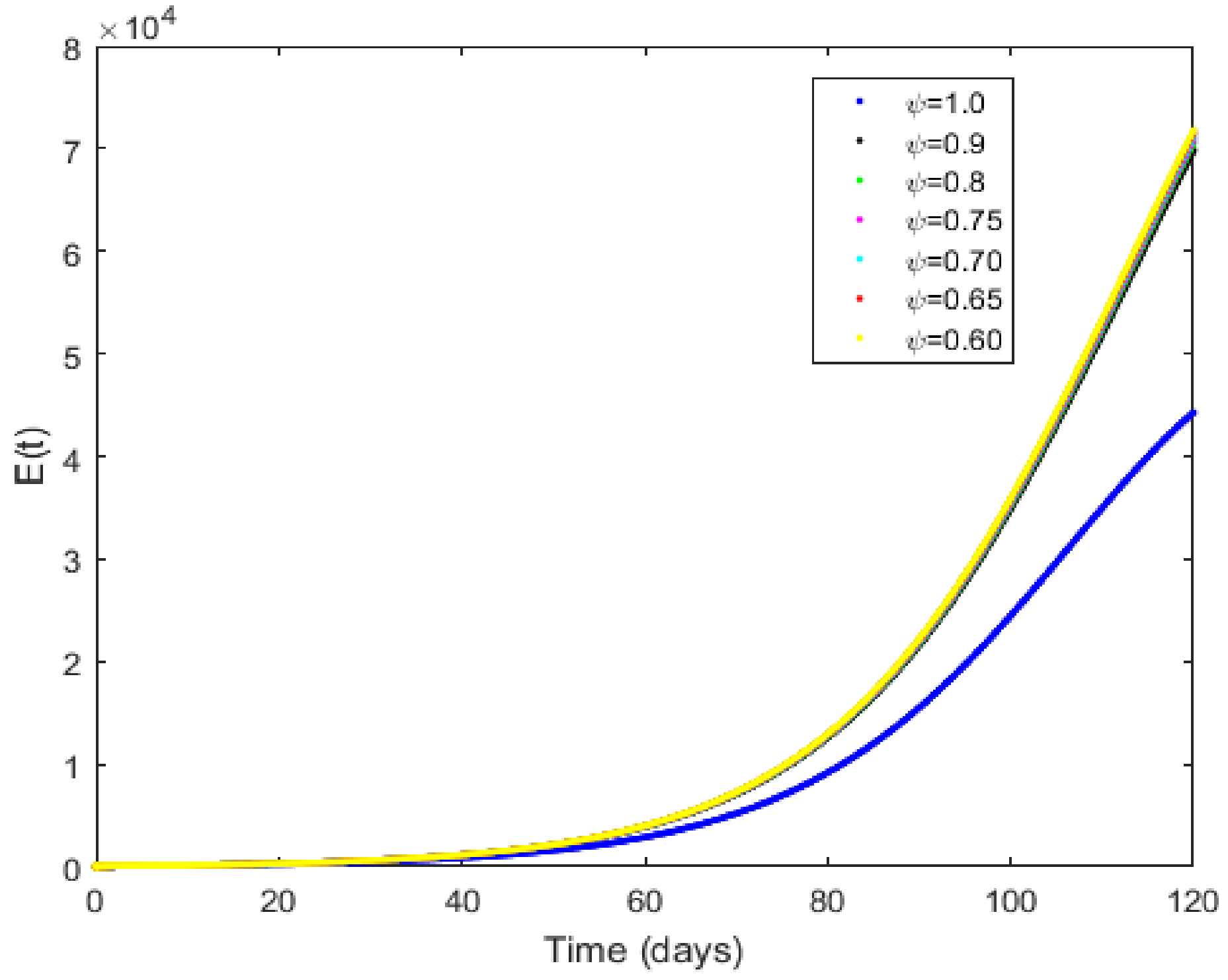

Fig.4: Behaviour of the exposed individuals given different values of $\psi$

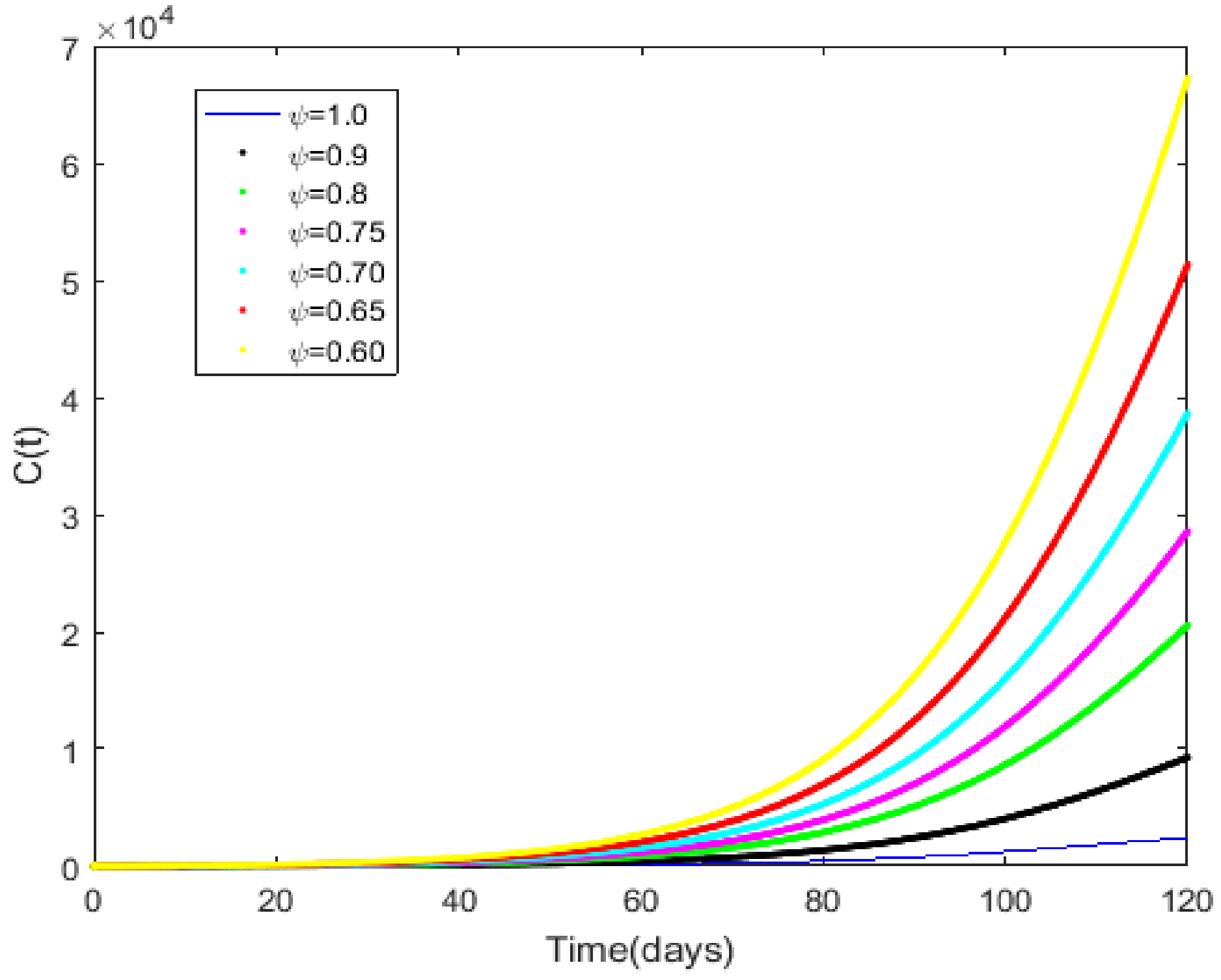

Fig.5: Behaviour of the COVID-19 infected individuals given different values of $\psi$

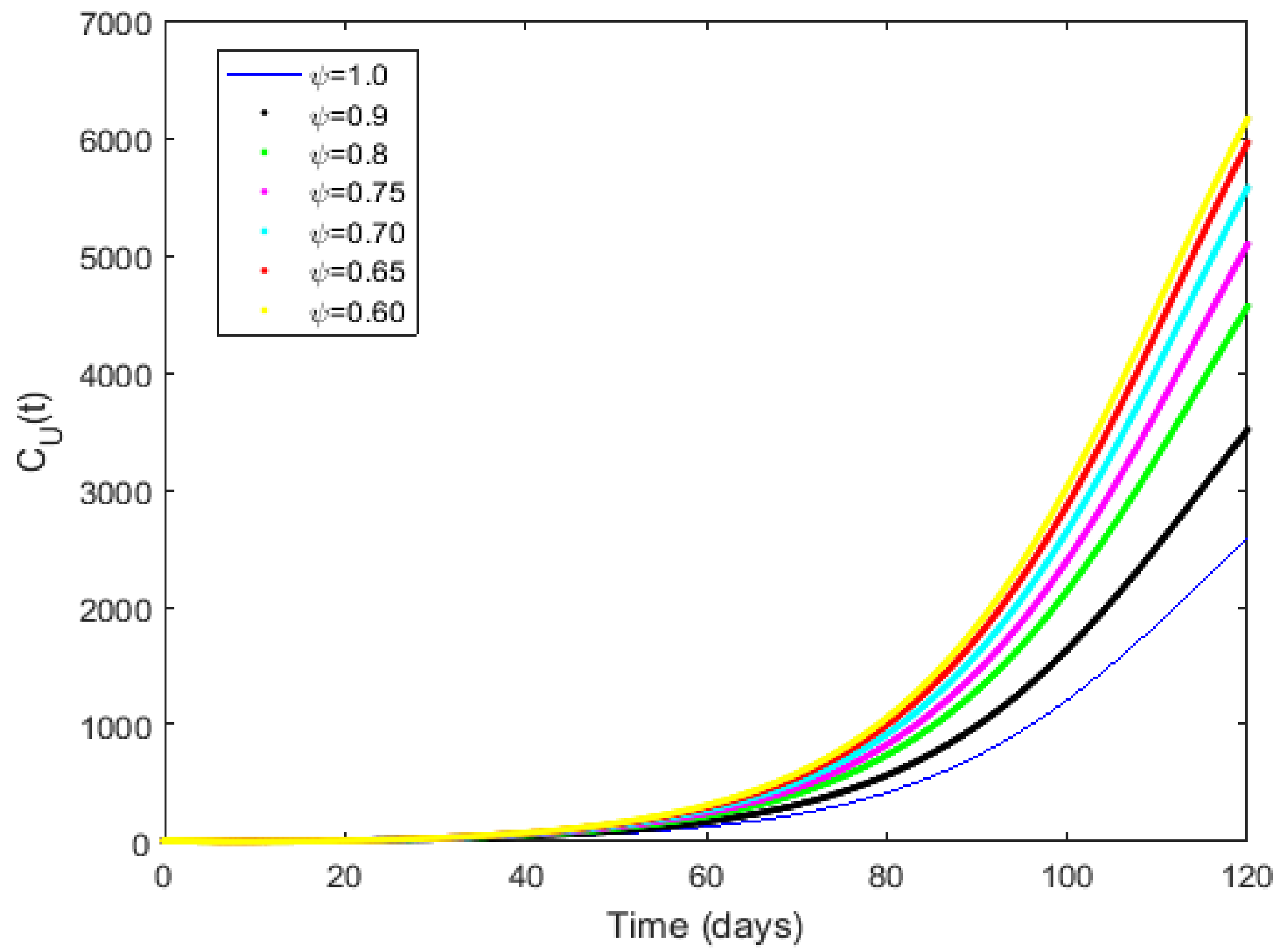

Fig.6: Behaviour of the COVID-19 infected individuals with an underlying condition given different values of $\psi$

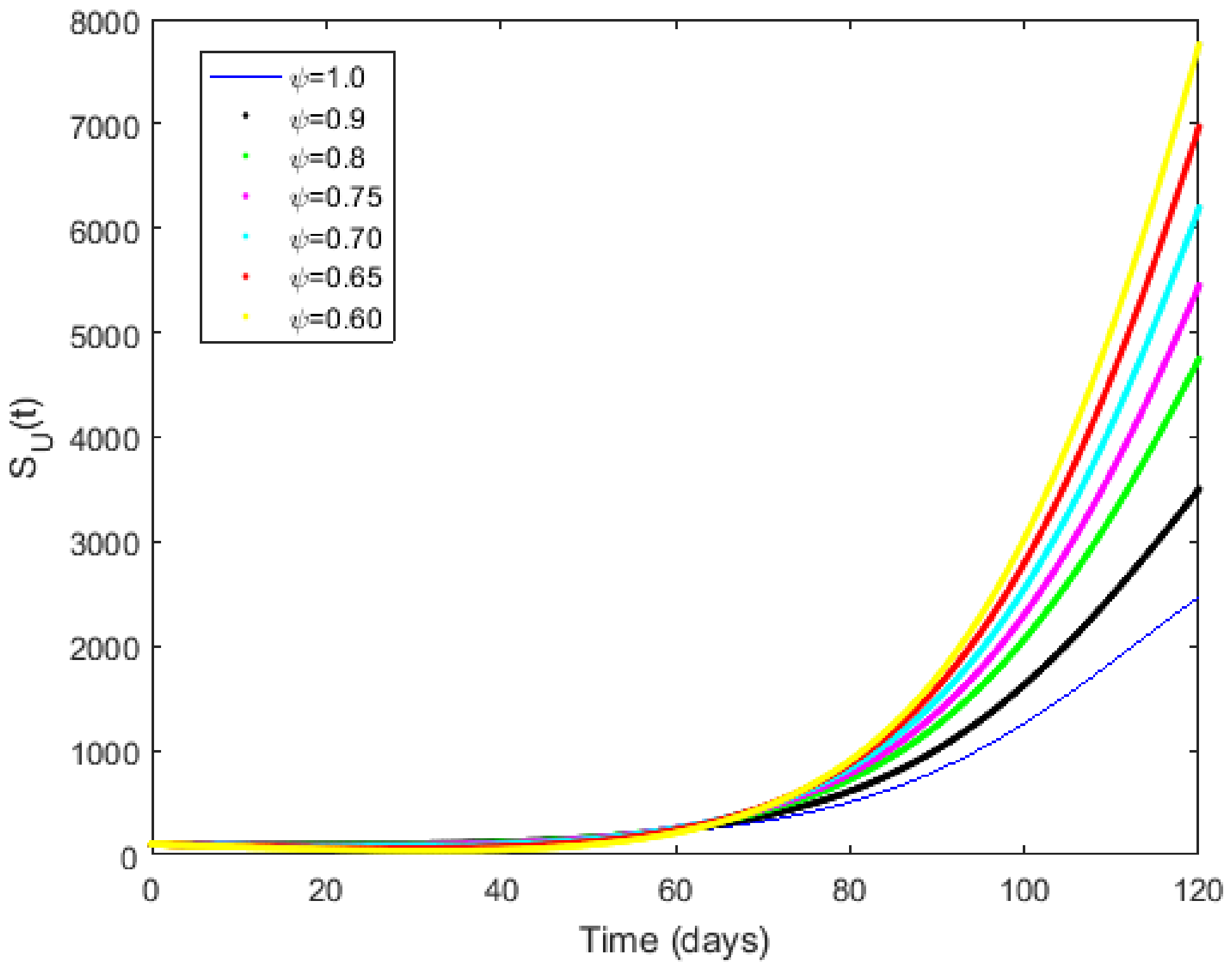

Fig.7: Behaviour of the susceptible individuals with an underlying condition given different values of $\psi$

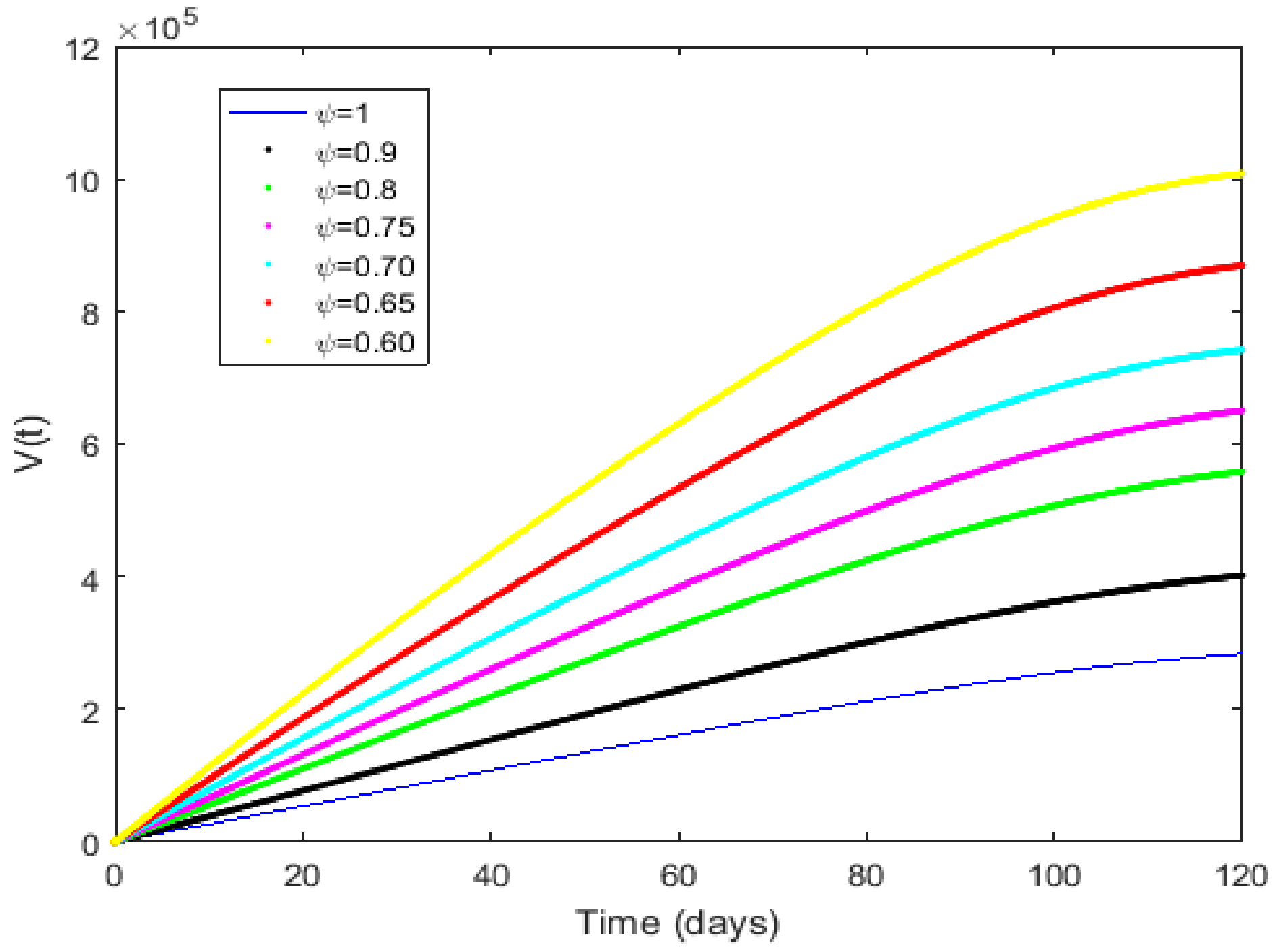

Fig.8: Behaviour of the vaccinated individuals given different values of $\psi$

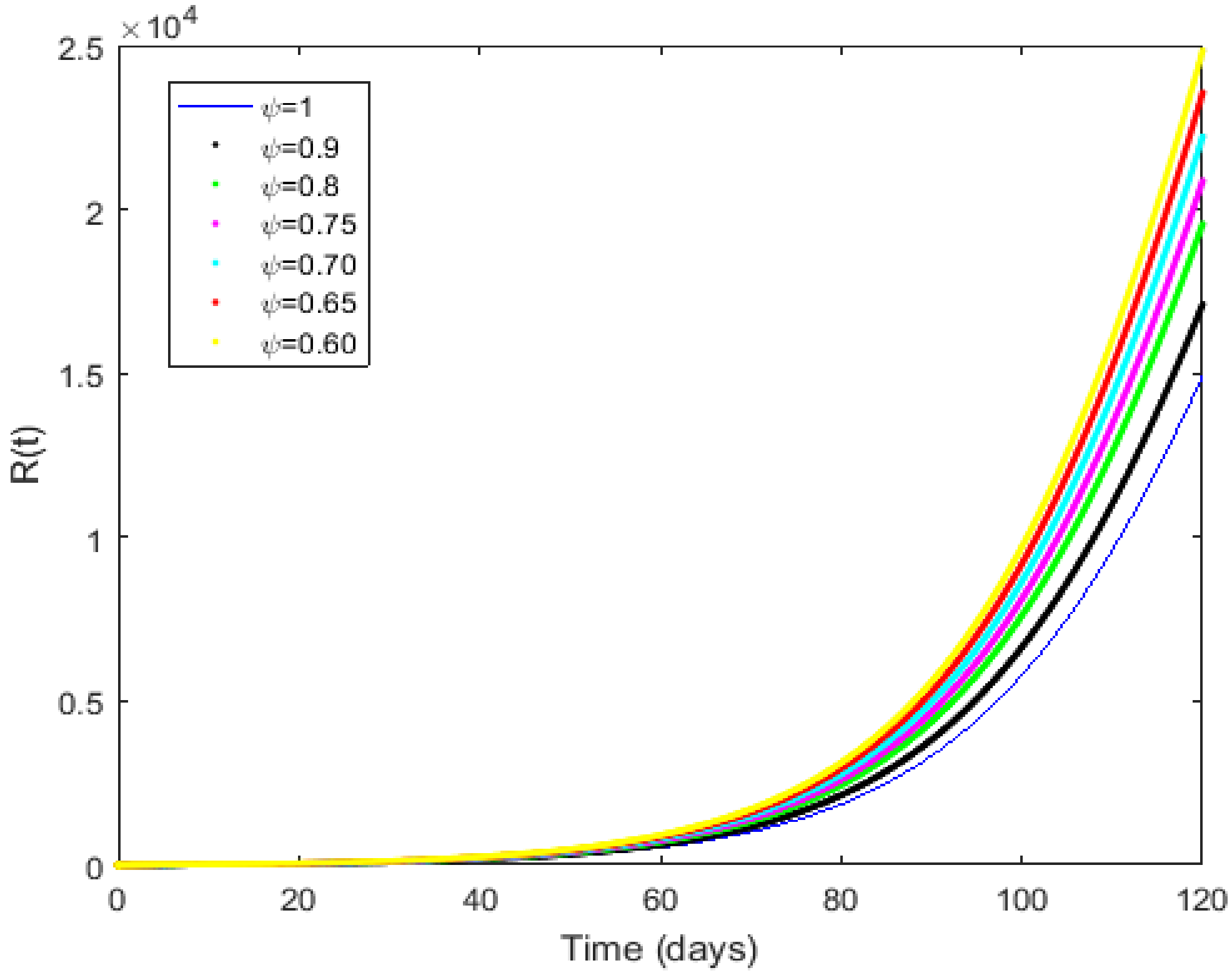

Fig.9: Behaviour of the recovered individuals given different values of $\psi$

Fig. 3 depicts the behaviour of the individuals susceptible to the COVID-19 disease for integer and non-integer values of the fractional operator $\psi$. The number of susceptible humans decreases as the fractional operator reduces from 1.0 within 120 days. Figure 4 depicts the behaviour of the exposed individuals for integer and non-integer values of the fractional operator $\psi$. The exposed population decreases as the fractional operator decreases from 0.60 within 120 days. Figure 5 depicts the behaviour of the COVID-19 patients without underlying conditions for integer and non-integer values of the fractional operator $\psi$. The number of COVID-19 patients without underlying condition reduces as the fractional operator increases from 0.60 within the 120 days. Figure 6 depicts the behaviour of the COVID-19 patients with an underlying condition for integer and non-integer values of the fractional operator $\psi$. The number of COVID-19 patients with an underlying condition decreases as the fractional operator increases from 0.60 within 120 days. Figure 7 depicts the behaviour of the susceptible with an underlying condition for integer and non-integer values of the fractional operator $\psi$. The number of susceptible with an underlying condition decreases as the fractional operator increases from 0.60. Figure 8 depicts the behaviour of the vaccinated individuals for integer and non-integer values of the fractional operator $\psi$. The number of vaccinated individuals decreases as the fractional-order operator increases from 0.60 within 120 days. Figure 9 depicts the behaviour of individuals removed from COVID-19. The number of recovered individuals decreases as the fractional operator increases from 0.60 within the 120 days. Hence to increase the number of individuals removed from the COVID-19, you need to reduce the fractional operator. Al so to reduce the total infections in both the population with and without an underlying condition, it requires an increment in the fractional operator.

## 7 Optimal Control Model

Let us consider the state system presented in system (1) with the control $u(t)$, where $u(t)$ represents vaccination of susceptible individuals with and without underlying conditions. We include the time-dependent control into the system (1) and we have

$$
\begin{aligned}
&{}^{ABC}D_{0,t}^{\psi}[S(t)] = \Omega^{\psi} - \beta^{\psi}\left(\frac{C+C_U}{N}\right)S - (\mu^{\psi} + \lambda^{\psi})S - uS, \\
&{}^{ABC}D_{0,t}^{\psi}[S_U(t)] = \lambda^{\psi}S + \gamma_1^{\psi}C_U - \beta^{\psi}\left(\frac{C+C_U}{N}\right)S_U - (\delta_2^{\psi} + \mu^{\psi})S_U - uS, \\
&{}^{ABC}D_{0,t}^{\psi}[E(t)] = \beta^{\psi}\left(\frac{C+C_U}{N}\right)S + \beta^{\psi}\left(\frac{C+C_U}{N}\right)S_U - (\varphi^{\psi} + \mu^{\psi})E, \qquad (28) \\
&{}^{ABC}D_{0,t}^{\psi}[C(t)] = \alpha\varphi^{\psi}E - (\delta^{\psi} + \mu^{\psi} + \gamma^{\psi})C, \\
&{}^{ABC}D_{0,t}^{\psi}[C_U(t)] = (1-\alpha)\varphi^{\psi}E - (\delta_1^{\psi} + \mu^{\psi} + \gamma_1^{\psi})C_U, \\
&{}^{ABC}D_{0,t}^{\psi}[V(t)] = \eta^{\psi}(S + S_U) - \mu^{\psi}V + u(S + S_U), \\
&{}^{ABC}D_{0,t}^{\psi}[R(t)] = \gamma^{\psi}C - \mu^{\psi}R.
\end{aligned}
$$

### 7.1 Analysis of the Optimal Control Model

We analyze the behaviour of the system (28). The objective function for fixed time $t_f$ is given by

$$
J(u_1,u_2) = \int_0^{t_f}[\, f_1S(t) + f_2S_U(t) + f_3E(t) + f_4C(t) + f_5C_U(t) + f_6V(t) + f_7R(t) + \frac{1}{2}Tu^2]dt. \quad (29)
$$

Where $f_1, f_2, ......, f_7$ are the relative weights and $T$ is the relative cost associated with the control $u$. The final time of the control is $t_f$. The control aims to minimize the cost function.

$$
J(u^*) = \min_{u_1,u_2 \in U} J(u). \qquad (30)
$$

Subject to the constraint system (28), where $0 \le u \le 1$ and $t \in (0,t_f)$. In other to derive the necessary condition for the optimal control, Pontryagin maximum principle given in [22] was used. This principle converts system (28) - (30) into a problem of minimizing a Hamiltonian H, defined by

$$\begin{aligned}
H = {} & f_1 S(t) + f_2 S_U(t) + \frac{1}{2}Tu^2 \\
& + \Lambda_S\{(\Omega^{\psi} - \beta^{\psi}\left(\frac{C+C_U}{N}\right)S - (\mu^{\psi} + \lambda^{\psi})S - uS\} \\
& + \Lambda_{S_U}\{\lambda^{\psi} S + \gamma_1^{\psi} C_U - \beta^{\psi}\left(\frac{C+C_U}{N}\right)S_U - (\delta_2^{\psi} + \mu^{\psi} + u)S_U\} \\
& + \Lambda_E\{\beta^{\psi}\left(\frac{C+C_U}{N}\right)S + \beta^{\psi}\left(\frac{C+C_U}{N}\right)S_U - (\varphi^{\psi} + \mu^{\psi})E\} \\
& + \Lambda_C\{\alpha\varphi^{\psi} E - (\delta^{\psi} + \mu^{\psi} + \gamma^{\psi})C\} \\
& + \Lambda_{C_U}\{(1-\alpha)\varphi^{\psi} E - (\delta_1^{\psi} + \mu^{\psi})C_U\} \\
& + \Lambda_V\{\eta^{\psi}(S + S_U) - \mu^{\psi} V + u(S + S_U)\} \\
& + \Lambda_R\{\gamma^{\psi} C - \mu^{\psi} R\}.
\end{aligned} \tag{31}$$

Where $\Lambda_S, \Lambda_{S_U}, \Lambda_E, \Lambda_C, \Lambda_{C_U}, \Lambda_V$ *and* $\Lambda_R$ represents the adjoint variables or co-state variables. The system of equations is derived by taking into account the correct partial derivatives of system (31) concerning the associated state variables.

**Theorem 4**: Given an optimal control $u^*(t)$ and corresponding solution $S^*, S_U{}^*, E^*, C^*, C_U{}^*, V^*, R^*$ that minimizes $J(u)$ over U. Furthermore, there exist adjoint variables $\Lambda_S, \Lambda_{S_U}, \Lambda_E, \Lambda_C, \Lambda_{C_U}, \Lambda_V, \Lambda_R$, satisfying

$$\begin{aligned}
& {}_t^{ABC}D_{t_f}^{\psi}\lambda_S = -\frac{\partial H}{\partial S}, {}_t^{ABC}D_{t_f}^{\psi}\lambda_{S_U} = -\frac{\partial H}{\partial S_U}, {}_t^{ABC}D_{t_f}^{\psi}\lambda_E = -\frac{\partial H}{\partial E}, {}_t^{ABC}D_{t_f}^{\psi}\lambda_C = -\frac{\partial H}{\partial C}, \\
& {}_t^{ABC}D_{t_f}^{\psi}\lambda_{C_U} = -\frac{\partial H}{\partial C_U}, {}_t^{ABC}D_{t_f}^{\psi}\lambda_V = -\frac{\partial H}{\partial V}, {}_t^{ABC}D_{t_f}^{\psi}\lambda_R = -\frac{\partial H}{\partial R},
\end{aligned} \tag{32}$$

$$0 = \frac{\partial H}{\partial u}, \tag{33}$$

$$
{}_0^{ABC}D_t^{\psi} S = -\frac{\partial H}{\partial \lambda_S}, {}_0^{ABC}D_t^{\psi} S_U = -\frac{\partial H}{\partial \lambda_{S_U}}, {}_0^{ABC}D_t^{\psi} E = -\frac{\partial H}{\partial \lambda_E}, {}_0^{ABC}D_t^{\psi} C = -\frac{\partial H}{\partial \lambda_C},
$$
$$
{}_0^{ABC}D_t^{\psi} C_U = -\frac{\partial H}{\partial \lambda_{C_U}}, {}_0^{ABC}D_t^{\psi} V = -\frac{\partial H}{\partial \lambda_V}, {}_0^{ABC}D_t^{\psi} R = -\frac{\partial H}{\partial \lambda_R}, \quad (34)
$$

With the transversality conditions

$$
\Lambda_S(t_f) = \Lambda_{S_U}(t_f) = \Lambda_E(t_f) = \Lambda_C(t_f) = \Lambda_{C_U}(t_f) = \Lambda_V(t_f) = \Lambda_R(t_f) = 0. \quad (35)
$$

***Proof:*** The differential equations characterized by the adjoint variables are obtained by considering the right-hand side differentiation of the system (31) determined by the optimal control. The adjoint equations derived are given as

$$
\begin{aligned}
\frac{d\Lambda_S}{dt} &= -f_1 + \beta^{\psi}\left(\frac{C + C_U}{N}\right)[\Lambda_S - \Lambda_E] + (\mu^{\psi} + u)\Lambda_S + \lambda^{\psi}[\Lambda_S - \Lambda_{S_U}] - (u + \eta^{\psi})\Lambda_V, \\
\frac{d\Lambda_{S_U}}{dt} &= -f_2 + \beta^{\psi}\left(\frac{C + C_U}{N}\right)[\Lambda_{S_U} - \Lambda_E] + (\delta_2^{\psi} + \mu^{\psi} + u)\Lambda_D - (u + \eta^{\psi})\Lambda_V, \\
\frac{d\Lambda_E}{dt} &= -f_3 + (\varphi^{\psi} + \mu^{\psi})\Lambda_E + \alpha\varphi^{\psi}[\Lambda_C - \Lambda_{C_U}] - \varphi^{\psi}\Lambda_{C_U}, \\
\frac{d\Lambda_C}{dt} &= -f_4 + \frac{\beta^{\psi} S}{N}[\Lambda_S - \Lambda_E] + \beta^{\psi}\frac{S_U}{N}[\Lambda_{S_U} - \Lambda_E] + (\gamma^{\psi} + \mu^{\psi} + \delta^{\psi})\Lambda_C - \gamma^{\psi}\Lambda_R, \quad (36) \\
\frac{d\Lambda_{C_U}}{dt} &= \frac{\beta^{\psi} S}{N}[\Lambda_S - \Lambda_E] + \beta^{\psi}\frac{S_U}{N}[\Lambda_{S_U} - \Lambda_E] - \gamma^{\psi}\Lambda_{S_U} + (\mu^{\psi} + \delta_1^{\psi})\Lambda_{C_U}, \\
\frac{d\Lambda_V}{dt} &= \mu^{\psi}\Lambda_V, \\
\frac{d\Lambda_R}{dt} &= \mu^{\psi}\Lambda_R.
\end{aligned}
$$

By obtaining the solution for $u^*$ subject to the constraints, we have

$$
0 = \frac{\partial H}{\partial u} = -Tu + S\Lambda_S + S_U\Lambda_{S_U} - (S + S_U)\Lambda_V. \quad (37)
$$

This gives

$$u^* = \min\left(1, \max\left(\frac{S\Lambda_S + S_U\Lambda_{S_U} - (S+S_U)\Lambda_V]}{T}\right)\right). \quad (38)$$

## 8 Numerical Techniques for the Fractional Optimal Control

Let us consider the following general initial value problem:

$${}_{\psi}^{ABC}D^{\psi} f(t) = g(t, f(t)), f(0) = f_0. \quad (39)$$

Applying the fundamental theorem of fractional calculus to equation (39), we obtain

$$f(t) - f(0) = \frac{1-\psi}{B(\psi)} g(t, f(t)) + \frac{\psi}{\Gamma(\psi)B(\psi)}\int_0^t g(\theta, f(\theta))(t-\theta)^{\psi-1} d\theta, \quad (40)$$

where $B(\psi) = 1-\psi+\frac{\psi}{\Gamma(\psi)B(\psi)}$ is a normalised function, and at $t_{n+1}$, we have

$$f_{n+1} - f_0 = \frac{\Gamma(\psi)(1-\psi)}{\Gamma(\psi)(1-\psi)+\psi} g(t_n, f(t_n)) + \frac{\psi}{\Gamma(\psi)+\psi(1-\Gamma(\psi))}\sum_{m=0}^{n}\int_{t_m}^{t_{n+1}} g\cdot(t_{n+1}-\theta)^{\psi-1} d\theta, \quad (41)$$

Now, approximating $g(\theta, f(\theta))$ in the interval $[t_k, t_{k+1}]$ using a two-step Lagrange interpolation given as follows [15, 42];

$$P = \frac{g(t_m, f_m)}{h}(\theta - t_{m-1}) - \frac{g(t_{m-1}, f_{m-1})}{h}(\theta - t_m) \quad (42)$$

Equation (42) is replaced in equation (41), and by performing the same steps in [15, 42] we obtain

$$f_{n+1} - f_0 = \frac{\Gamma(\psi)(1-\psi)}{\Gamma(\psi)(1-\psi)+\psi} g(t_n, f(t_n)) + \frac{1}{(\psi+1)+(1-\psi)\Gamma(\psi)+\psi}$$

$$\sum_{m=0}^{n} h^{\psi} g(t_m, f(t_m))(n+1-m)^{\psi}(n-m+2+\psi)-(n-m)^{\psi}(n-m+2+2\psi)-h^{\psi} g(t_{m-1}, \quad (43)$$

$$f(t_{m-1}))(n+1-m)^{\psi+1}(n-m+2+\psi)-(n-m+1+\psi).$$

To obtain high stability, we present a simple modification in equation (43). This modification is to replace the step size h with $\phi(h)$

$$\phi(h) = h + O(h^2), \quad 0 \prec \phi(h) \le 1$$

The new scheme is called the nonstandard two-step Lagrange interpolation method (NS2LIM) and is given as:

$$f_{n+1} - f_0 = \frac{\Gamma(\psi)(1-\psi)}{\Gamma(\psi)(1-\psi)+\psi} g(t_n, f(t_n)) + \frac{1}{(\psi+1)+(1-\psi)\Gamma(\psi)+\psi}$$

$$\sum_{m=0}^{n} \phi(h)^{\psi} g(t_m, f(t_m))(n+1-m)^{\psi}(n-m+2+\psi)-(n-m)^{\psi}(n-m+2+2\psi)-\phi(h)^{\psi} g(t_{m-1}, f(t_{m-1})) \quad (44)$$

$$(n+1-m)^{\psi+1}(n-m+2+\psi)-(n-m+1+\psi).$$

## 9 Numerical Simulation of the Optimal Control Model

In this section, we discuss numerically the behaviour of the optimal control system (28) and show the results of varying the control ( $u_1$ ). By using the parameter values given in Table 1 and the same initial conditions, the results of the simulations are displayed in Figs. 10 – 14

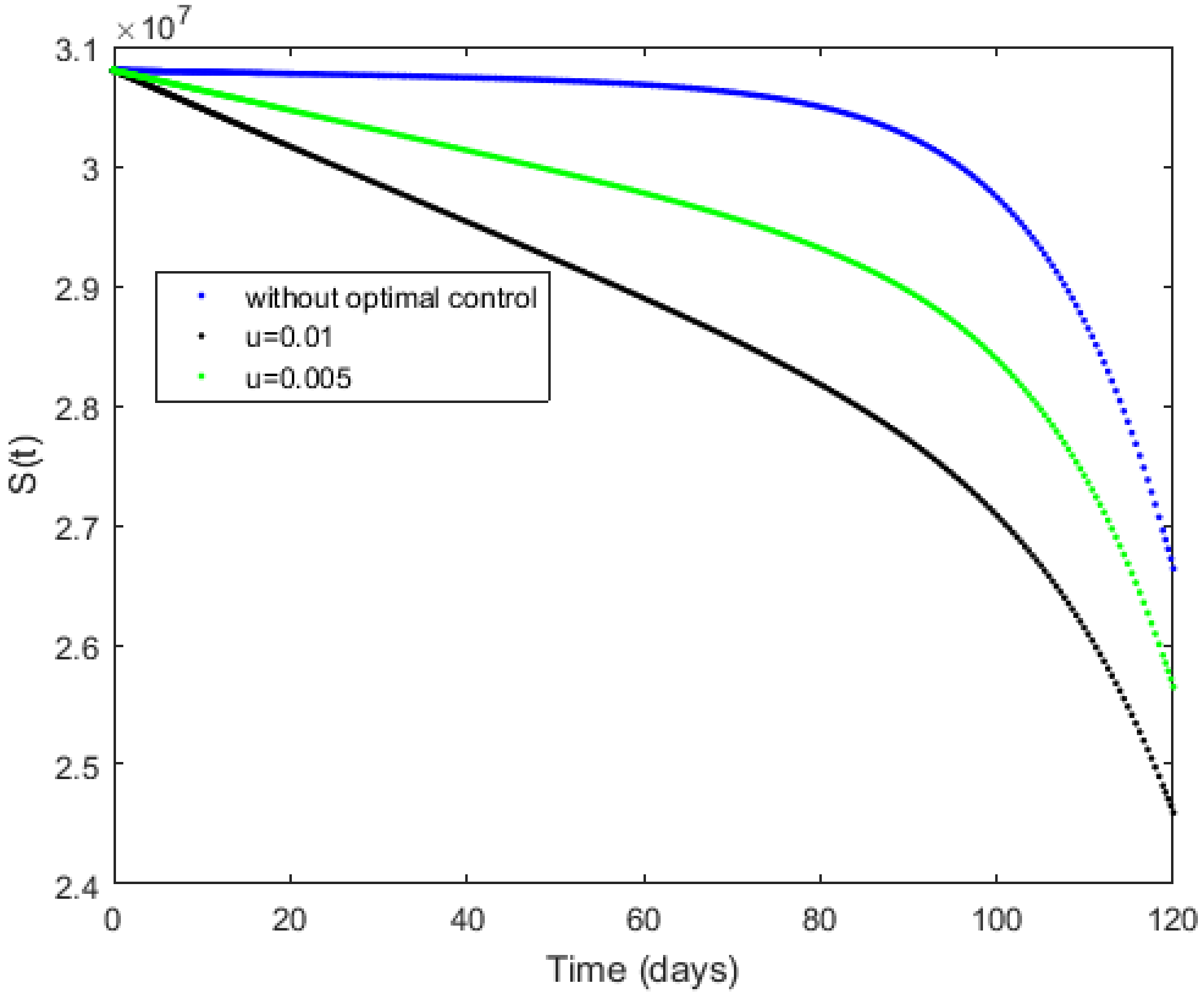

Fig. 10: Behaviour of the susceptible population with and without control

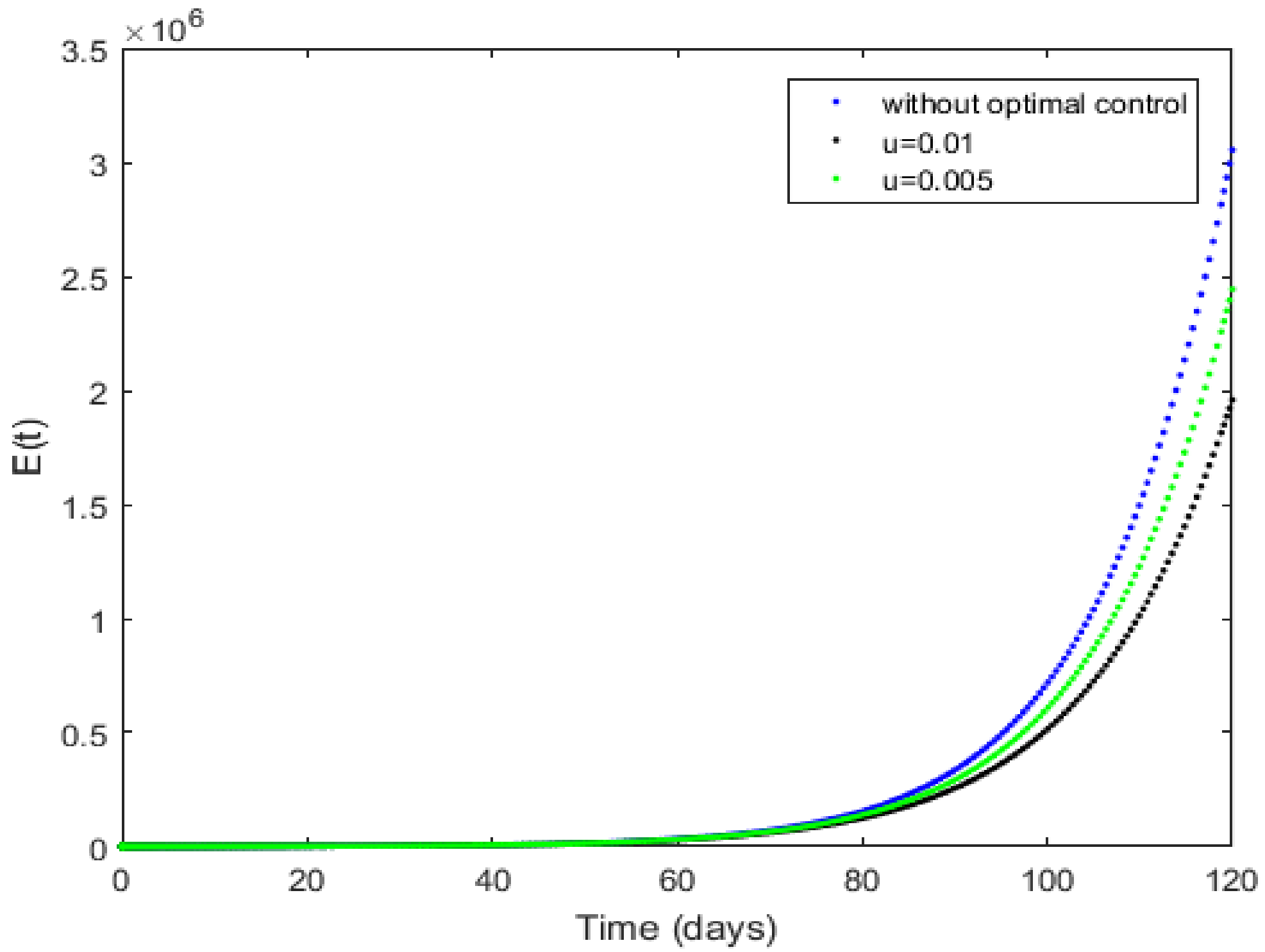

Fig. 11: Behaviour of the exposed individuals with and without control

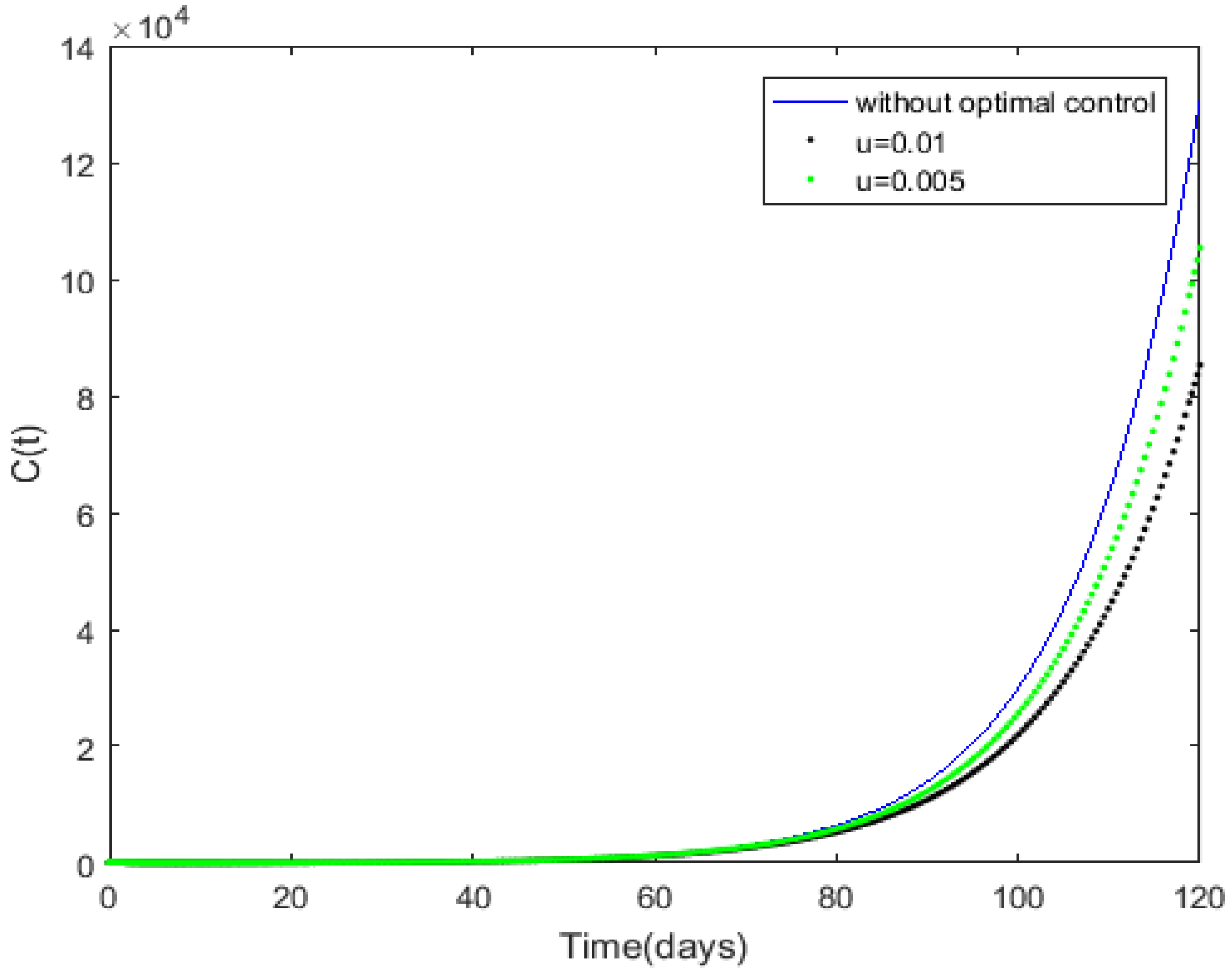

Fig. 12: Behaviour of the COVID-19 patients with no underlying conditions with and without control

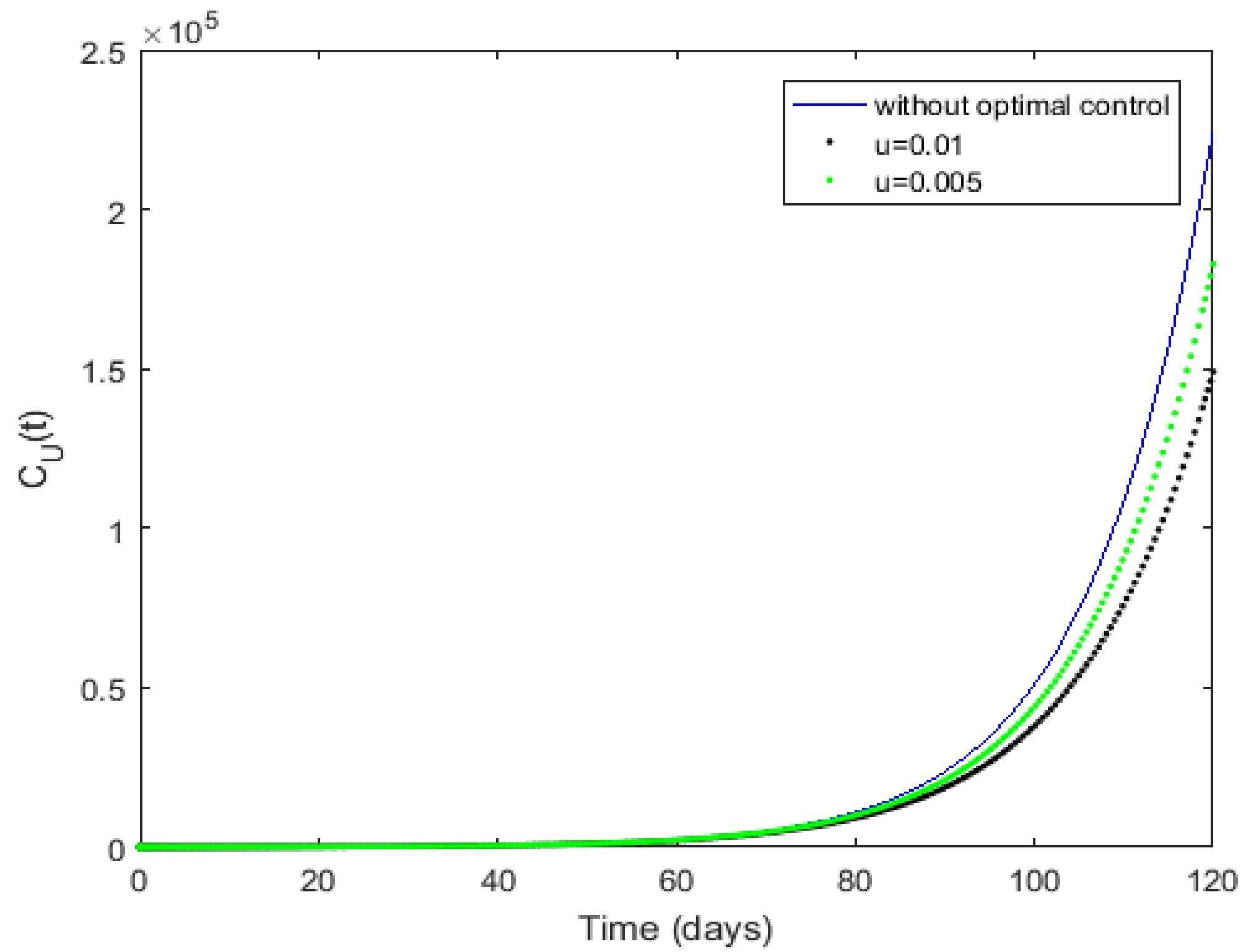

Fig. 13: Behaviour of the COVID-19 patients with underlying condition with and without control

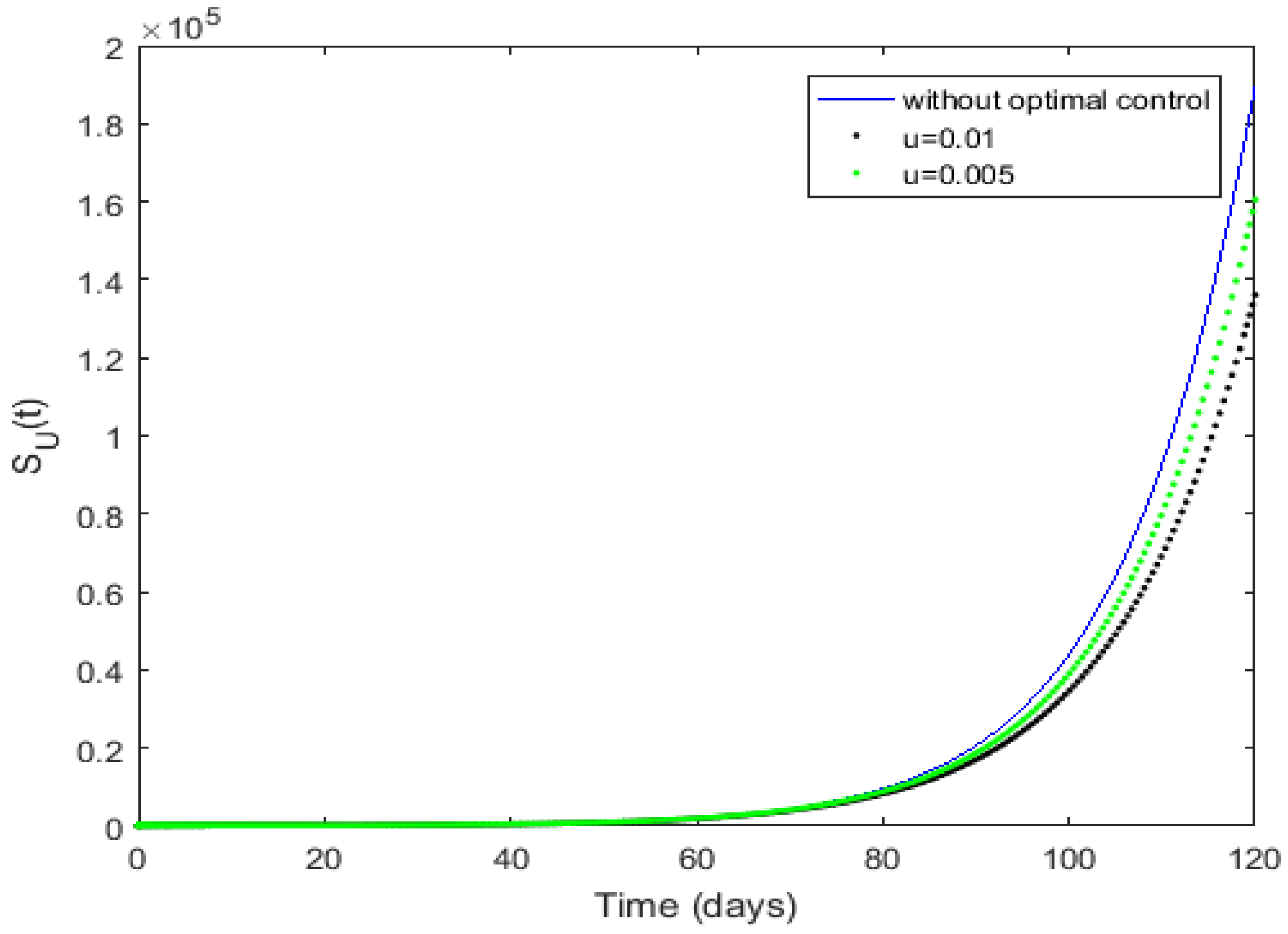

Fig. 14: Behaviour of the susceptible with an underlying condition with and without optimal control

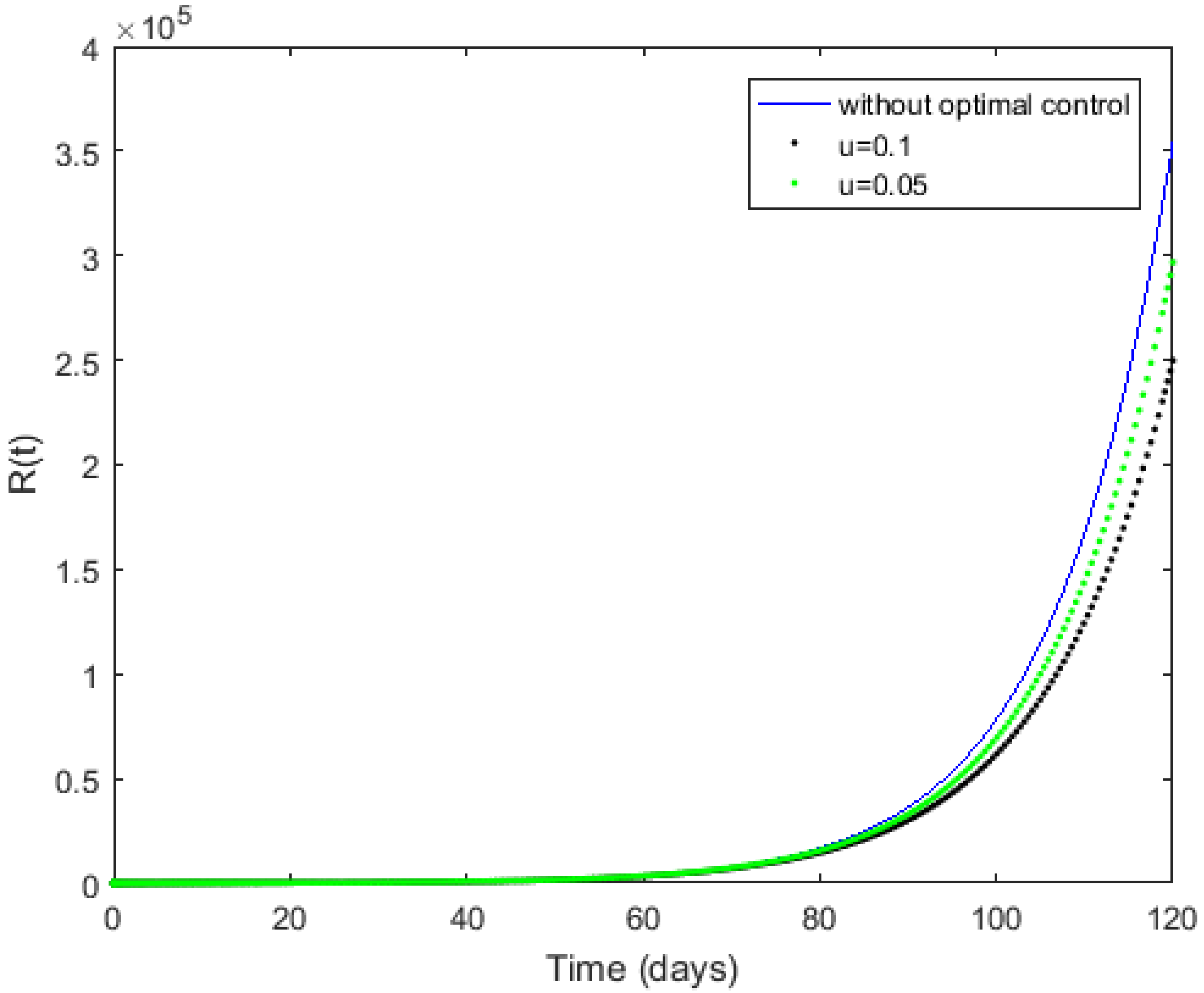

Fig. 15: Behaviour of the recovered individuals with and without control

Figure 10 depicts the behaviour of the susceptible population with and without optimal control. The number of susceptible decreases as the optimal control parameter u(t) increases from 0.0005 to 0.001. This shows as more and more people vaccinate, people develop immunity to the COVID-19. Figure 11 depicts the behaviour of the exposed individuals with and without optimal control. The number of people who get exposed to the COVID-19 decreases as the optimal control increases from 0.0005 to 0.001 within 120 days. Figure 12 depicts the behaviour of the COVID-19 patients without an underlying condition. The number reduces as more people vaccinate against the COVID-19. In figure 13, the number of COVID-19 patients with an underlying condition decreases when there is a vaccination control. In figure 14, vaccinating the susceptible population with an underlying condition leads to a decrease in the number of susceptible with an underlying condition. Figure 15 depicts the behaviour of the individuals removed from the COVID-19. An intense vaccination leads to a decline in the recovered population as fewer individuals contract the disease.

## 10 Conclusion

In this study, a COVID-19 model taking into consideration a population with an underlying condition has been examined using the fractional-order derivative defined in the Atangana-Beleanu and Caputo sense. The qualitative properties of the model were examined. The basic reproductive number of the model was determined. The equilibrium points of the model were found and stability analyses were carried out. The endemic equilibrium was locally stable for $R_o > 1$. The existence and the uniqueness of the solution are established along with Hyers –Ulam Stability. The numerical scheme for the operator was carried out to obtain a numerical simulation to support the analytical results. The simulation reveals a decline in infections as the fractional operator was increased from 0.6 within the 120 days. The numerical simulation of the optimal control revealed vaccination reduces the number of individuals susceptible to the COVID-19, individuals exposed to the COVID-19, Covid-19 patients without an underlying condition and those with an underlying condition.

## Data Availability

The data/information supporting the formulation of the mathematical model in this paper are/is from the Ghana health service website: https://www.ghs.gov.gh/covid19/ which has been cited in the manuscript.

## Declaration of interest

No conflict of interest regarding the content of this article

## Funding

The research did not receive funding from any sources

## Acknowledgement

This manuscript was submitted as a pre-print in the link https://arxiv.org/ftp/arxiv/papers/2201/2201.11659.pdf and has been referenced.